\documentclass[10pt,journal,compsoc]{IEEEtran}
\IEEEoverridecommandlockouts

\usepackage[pdftex]{graphicx}
\usepackage{color}

\usepackage{amssymb,amsmath}
\usepackage{proof,latexsym}
\usepackage{color}
\usepackage{enumerate}

\usepackage{ifthen}
\usepackage{amssymb}

\usepackage{float}
\usepackage{alltt}
\usepackage{color}
\usepackage{graphicx}
\usepackage{multirow}
\usepackage{epstopdf}
\usepackage{hyperref}
\usepackage{url}
\usepackage{graphicx}
\usepackage{epstopdf}
\usepackage{balance}
\usepackage{tabularx}
\usepackage{array}
\usepackage{color}

\usepackage{amssymb,amsmath}
\usepackage{proof,latexsym}
\usepackage{color}
\usepackage{enumerate}

\usepackage{ifthen}
\usepackage{amssymb}

\usepackage{float}
\usepackage{alltt}
\usepackage{color}
\usepackage{graphicx}
\usepackage{multirow}
\usepackage{epstopdf}
\usepackage{hyperref}
\usepackage{url}
\usepackage{algorithm}
\usepackage{algorithmicx}
\usepackage{epstopdf}
\usepackage{balance}
\usepackage{multirow}
\usepackage{algpseudocode}
\usepackage{booktabs}
\usepackage{graphicx}
\usepackage{subcaption}
\usepackage{adjustbox}
\usepackage{array}
\usepackage{tabularx}

\usepackage{xcolor}
\newcommand\recheck[1]{\textcolor{black}{#1}}
\newcommand\recheckagain[1]{\textcolor{black}{#1}}

\newcommand\rechecknewagain[1]{\textcolor{black}{#1}}

\algrenewcommand\algorithmicrequire{\textbf{Inputs:}}
\algrenewcommand\algorithmicensure{\textbf{Outputs:}}

\newboolean{showcomments}
\setboolean{showcomments}{true}
\ifthenelse{\boolean{showcomments}}
{ \newcommand{\mynote}[2]{
    \fbox{\bfseries\sffamily\scriptsize#1}
    {\small$\blacktriangleright$\textsf{\emph{#2}}$\blacktriangleleft$}}}
{ \newcommand{\mynote}[2]{}}

\begin{document}
\bstctlcite{IEEEexample:BSTcontrol}
\pdfpagewidth 8.5in
\pdfpageheight 11.0in

\title{
Network-Clustered Multi-Modal Bug Localization}


\author{\IEEEauthorblockN{Thong Hoang, Richard J. Oentaryo, Tien-Duy B. Le, and David Lo}\\
\IEEEauthorblockA{School of Information Systems, Singapore Management University\\
\{vdthoang.2016, roentaryo, btdle.2012, davidlo\}@smu.edu.sg}
}

\maketitle
\begin{abstract}
Developers often spend much effort and resources to debug a program. To help the developers debug, numerous information retrieval (IR)-based and spectrum-based bug localization techniques have been devised. IR-based techniques process textual information in bug reports, while spectrum-based techniques process program spectra (i.e., a record of which program elements are executed for each test case). While both techniques ultimately generate a ranked list of program elements that likely contain a bug, they only consider one source of information---either bug reports or program spectra---which is not optimal. In light of this deficiency, this paper presents a new approach dubbed \underline{Net}work-clustered \underline{M}ulti-modal Bug \underline{L}ocalization (NetML), which utilizes multi-modal information from both bug reports and program spectra to localize bugs. NetML facilitates an effective bug localization by carrying out a joint optimization of bug localization error and clustering of both bug reports and program elements (i.e., methods). The clustering is achieved through the incorporation of \emph{network Lasso} regularization, which incentivizes the model parameters of similar bug reports and similar program elements to be close together.  To estimate the model parameters of both bug reports and methods, NetML employs an adaptive learning procedure based on Newton method that updates the parameters on a per-feature basis. \recheck{Extensive experiments on 355 real bugs from seven software systems have been conducted to benchmark NetML against various state-of-the-art localization methods. The results show that NetML surpasses the best-performing baseline by 31.82\%, 22.35\%, 19.72\%, and 19.24\%, in terms of the number of bugs successfully localized when a developer inspects the top 1, 5, and 10 methods and Mean Average Precision (MAP), respectively.}

\end{abstract}

\section{Introduction}
\label{sec.intro}

\recheck{Debugging bug reports, which often come in high volume~\cite{AnvikHM05}, has proved to be a difficult task that takes much resources and time~\cite{tassey02economic}. Various techniques have been devised to help developers locate buggy program elements from their symptoms. These symptoms could be in the form of description of a bug experienced by a user, or a failing test case. These techniques---often collectively referred to as bug (or fault) localization---analyze the symptoms of a bug and produce a list of program elements ranked based on their likelihood to contain the bug. In general, a program element can be defined at three levels of granularity, i.e., source file level, method/function level, and line of code level.}

\subsection{The Need for Multi-modal Bug Localization}
\label{sec:need_buglocalization}

Existing bug localization techniques broadly fall into two major categories: \emph{information retrieval} (IR)-based techniques~\cite{Rao:2011:RSL:1985441.1985451,Sisman:2012:IVH:2664446.2664454,Zhou:2012:BFM:2337223.2337226,SahaLKP13}, and \emph{spectrum}-based bug localization techniques~\cite{JH05,Abreu:2009,RR03,ZH02,Zeller2002a,CZ05,LAZJ03,Libl+05,LYFHM05}. The IR-based bug localization techniques typically analyze textual descriptions contained in bug reports and identifier names and comments in source code files. They then return a ranked list of program elements (typically program files) that are the most similar to the bug textual description. The spectrum-based bug localization techniques typically analyze program spectra that corresponds to program elements that are executed by failing and successful execution traces. Likewise, they return a ranked list of program elements (typically program blocks or statements) that are executed more often in the failing rather than correct traces.


The above-mentioned approaches, however, only consider one kind of symptom or one source of information, i.e., only bug reports or only execution traces. This is a limiting factor since hints of the location of a bug may be spread in both bug report and execution traces; and some hints may only appear in one but not the other. In this work, we put forward a bug localization approach that addresses the deficiency of existing methods by jointly utilizing both bug reports and execution traces. We refer to this approach as {\em multi-modal bug localization}, as we need to consider multiple modes of inputs (i.e., bug reports and program spectra). Such an approach fits well to developers' debugging activities as illustrated by the following scenarios:

\begin{enumerate}
	\item Developer $D$ is working on a bug report that is submitted to Bugzilla. One of the first tasks that he needs to do is to replicate the bug based on the description in the report. If the bug can be successfully replicated, he will proceed to the debugging step; otherwise, he will mark the bug report as ``WORKSFORME'' and will not continue further~\cite{worksforme}. After $D$ replicates the bug, he has one or a few failing execution traces. He also has a set of regression tests that he can run to get successful execution traces. Thus, after the replication process, $D$ has {\em both} the textual description of the bug and a program spectra that characterizes the bug. With this, $D$ can proceed to use multi-modal bug localization.
	\item  Developer $D$ runs a regression test suite and some test cases fail. Based on his experience, $D$ has some idea why the test cases fail. $D$ can create a textual document describing the bug. At the end of this step, $D$ has {\em both} program spectra and textual bug description, and can proceed to use multi-modal bug localization which will leverage not only the program spectra but also $D$'s domain knowledge to locate the bug.
\end{enumerate}

It is worth noting that our work focuses on localizing a bug to the \emph{method} that contains it. Historically, most IR-based bug localization techniques aim at finding buggy files~\cite{Rao:2011:RSL:1985441.1985451,Sisman:2012:IVH:2664446.2664454,Zhou:2012:BFM:2337223.2337226,SahaLKP13}, while most spectrum-based techniques find buggy lines~\cite{JH05,Abreu:2009,RR03}. \recheck{Localization at the method level can be a good tradeoff. That is, a method is not as big as a file, but it often contains sufficient context needed to help developers understand a bug. On the other hand, by just looking at a line of code, developers often cannot determine whether it is the location of the bug or understand the bug well enough to fix it~\cite{ParninO11}. Admittedly, if the methods are long, a finer granularity (e.g., basic blocks) may be preferred. Nevertheless, a recent study by Kochhar et al.~\cite{Kochhar:2016:PEA:2931037.2931051} highlights that out of the 386 practitioners they surveyed, the majority indicates method-level as the preferred granularity.}

\subsection{Proposed Approach}
\label{sec:proposed_intro}

In this paper, we present a new approach called the \underline{Net}work-clustered \underline{M}ulti-modal Bug \underline{L}ocalization (NetML), which works based on three main intuitions:

\begin{enumerate}
	\item \recheck{Firstly, it is recognized that a large variety of bugs exist, and different bugs need different treatments \cite{ThungWLJ12,XiaZLZ13}. A bug report written by a developer provides unique description of a bug. Thus, different bugs require separate model parameters to capture their individual characteristics. Similarly, different program elements (or methods in this work) are of different nature, and should be characterized by separate model parameters.}
	
	\item \recheck{A recent study by Parnin and Orso~\cite{ParninO11} also showed that some words are more useful in localizing bugs}, and suggested that ``future research could also investigate ways to automatically suggest or highlight terms that might be related to a failure''. Our NetML provides such capability by incorporating \emph{method suspiciousness} feature, which allows us to automatically highlight suspicious terms and use them to localize bugs.
	
	\item \recheck{We also observe that bugs and program elements are \emph{not} completely independent, and some bug reports (or methods) may be more similar to certain bug reports (or methods) than to others}. As such, similar bugs (or methods) should have model parameters that are close together. \recheck{This enforcement of clustering of model parameters would enable similar bug reports (or methods) to share information and reinforce one another.} 
\end{enumerate}


\recheck{The first two intuitions have been captured in our recent work---dubbed \underline{A}daptive \underline{M}ulti-Modal Bug \underline{L}ocalization (AML)~\cite{Le:2015:IRS:2786805.2786880}---which we extend in this paper. In particular, AML already incorporates the ideas of adaptively computing separate model parameters for each bug report, and of computing the method suspiciousness feature\footnote{\recheck{To understand the concept of feature and model parameters, we can draw an analogy to a linear model $y = \sum_i w_i x_i$. A feature refers to the (independent) input variable $x_i$, while a model parameter refers to the weight coefficient $w_i$ for each feature $x_i$. In this case, the model parameters $w_i$ need to be learned/estimated from data.}}.}
However, the current AML approach exhibits two main shortcomings. Firstly, AML only has the concept of model parameters for bug reports, but not for program elements (or methods). As such, it is not able to capture  variation in the inherent characteristics of different program elements (methods), which may limit its effectiveness in localizing a bug. Secondly, the model parameters of each bug report are learned independently of those of other bug reports. As a result, AML is unable to take advantage of the clustering/similarity traits of different bug reports in the localization process.

\begin{figure*}[!t]
	\centering
	\begin{minipage}[t]{.50\textwidth}
		\centering
		\begin{tabular}{|p{0.95\textwidth}|}
			\hline
			\textbf{Bug 30798}\\ 
			\hline
			\texttt{Summary:} 
			\textcolor{blue}{JUnit} shows \textcolor{magenta}{output} implementation grabs System.out and System.err later than it should.  
			\\
			\\
			\texttt{Description:}
			\\
			\textbf{What steps will reproduce the problem?} \textcolor{blue}{JUnit}\textcolor{red}{Test}Runner creates the
			\textcolor{blue}{junit}.framework.\textcolor{red}{Test} instance before grabbing System.out and System.err. As a
			result, anything printed to System.out or System.err in the constructor \textbf{\dots}
			\\
			\hline
		\end{tabular}
	\end{minipage}\hfill
	\begin{minipage}[t]{.50\textwidth}
		\centering
		\begin{tabular}{|p{0.95\textwidth}|}
			\hline
			\textbf{Bug 43969}\\ 
			\hline
			\texttt{Summary:} \textcolor{blue}{JUnit}4 \textcolor{red}{test}s marked @Ignore do not appear in XML \textcolor{magenta}{output} \\
			\\
			\texttt{Description:} 
			
			\textbf{What steps will reproduce the problem?} Run a \textcolor{blue}{JUnit} 4 \textcolor{red}{test} marked with the @Ignore annotation.  The test will not appear at all in the XML \textcolor{magenta}{output}.

			\\
			\hline
		\end{tabular}
	\end{minipage}
	\caption{Example of two bug reports which have the same faulty method in project Apache-Ant~\cite{ant_link}. The colored text indicates some common word tokens that these two bugs share.}
	\label{fig:bug_motivation}
\end{figure*}

The proposed NetML method addresses these shortcomings by performing joint optimization of localization loss function and clustering of both bug reports and methods. Specifically, it generalizes AML in two important ways:

\begin{enumerate}
	\item NetML provides a richer model that has two sets of (model) parameters---one for bug reports and the other for methods. The addition of the method parameters (in contrast to AML that has only bug report parameters) provides NetML with a higher degree of freedom to characterize the different variety of bug reports and methods more accurately.
	
	\item NetML incorporates network Lasso regularization~\cite{Hallac:2015:NLC:2783258.2783313} into its parameter learning procedure, which enforces similar bug reports (and methods) to have similar (or even identical) model parameters. This clustering enforcement would allow similar bug reports (or methods) to reach a consensus on the model parameters, leading to a simpler ``policy'' for bug localization. This enables the models of bug reports (or methods) to complement and borrow strength from one another. In turn, this would improve robustness and generalization performance on new/unseen bug reports.
\end{enumerate}

It is noteworthy that, deviating from the conventional network Lasso~\cite{Hallac:2015:NLC:2783258.2783313} which deals with only a single network (graph), we impose regularization over two networks, i.e., bug report similarity and method similarity graphs. This allows us to achieve simultaneous clustering of both bug reports and methods, and exploit their similarity traits so as to achieve a more effective bug localization.

To illustrate how the network Lasso regularization in NetML can benefit bug localization, Fig. ~\ref{fig:bug_motivation} shows two bug reports from Apache-Ant~\cite{ant_link} project, namely Bug 30798 and Bug 43969. These two bug reports describe issues with the ``showoutput'' option for Apache Ant's JUnit task and the corresponding bugs both reside in the \texttt{run} method of the \texttt{JUnitTestRunner.java} file. Bug 30798 mentions the names of a few source code files and one of them is the name of the buggy file (i.e., \texttt{JUnitTestRunner}), while no such hint is included in Bug 43969. AML manages to successfully localize the buggy method for Bug 30798, by ranking it high in the returned ranked list. However, due to the limited information in Bug 43969, it is not able to do the same for it. Upon a closer investigation, we can see that Bug 30798 and Bug 43969 are similar, since they share a number of common word tokens (i.e., ``JUnit'', ``text'', ``output'', etc.). The network Lasso regularization is able to take advantage of this similarity by enforcing similar bug reports to have similar model parameters. In such way, NetML leverages the similarity of Bug 30798 and Bug 43969 to guide/reinforce the prediction for Bug 43969, which leads to successful localization for both bugs.

\subsection{Contributions}
\label{sec:contributions}


\rechecknewagain{To evaluate the efficacy of the NetML approach, we conducted experiments using a dataset of 355 real bugs from seven medium to large software systems: Ant, AspectJ, Lang, Lucene, Math, Rhino, and Time. All real bug reports and real test cases were collected from these systems. The test cases were run to generate program spectra. We compare NetML with our previous AML method. Additionally, we evaluate our approach against a wide range of state-of-the-art approaches, including two multi-modal feature localization techniques (i.e., PROMESIR~\cite{PoshyvanykGMAR07}, DIT$^\text{A}$ and DIT$^\text{B}$~\cite{DitRP13}), four spectrum-based bug localization techniques (\cite{Abreu:2007:ASF:1308173.1308264, DBLP:journals/tr/WongDGL14, XuanM14, B.Le:2016:LBF:2931037.2931049}), and an IR-based bug localization technique (i.e., LR$^A$ and LR$^B$~\cite{YeBL14}). We use two well-known evaluation metrics to estimate the performance of our approach: number of bugs localized by inspecting the top N program elements (Top N) and mean average precision (MAP). 
Note that Top N and MAP have been widely used in past bug localization studies,
e.g.,~\cite{Rao:2011:RSL:1985441.1985451,Sisman:2012:IVH:2664446.2664454,Zhou:2012:BFM:2337223.2337226,SahaLKP13}.}

Our experiment results demonstrate that, among the 355 bugs, NetML can successfully localize 116, 219, and 255 bugs when developers only inspect the Top 1, Top 5, and Top 10 methods in the lists that NetML produces, respectively. These constitute 31.82\%, 22.35\%, 19.72\%, and 19.24\% improvements over AML (which is the second best method in our benchmark), in terms of Top 1, Top 5 , Top 10, and MAP results respectively.

We summarize the key contributions of this paper below:

\begin{enumerate}
\item We present a novel multi-modal bug localization method that adaptively learns two sets of model parameters that characterize each bug report and method, respectively. We are also the first to incorporate the network Lasso regularization on both bug report and method similarity networks, which facilitates an effective joint optimization of bug localization quality and clustering of both bug reports and methods.  
\item We develop an adaptive learning procedure based on Newton update to jointly update the model parameters of bug reports and methods on a per-feature basis. The procedure is based on the formulation of strict convex loss function, which provides a theoretical guarantee that any minimum found will be globally optimal. 
\item We have extensively evaluated NetML on a dataset of 355 real bugs from seven software systems using real bug reports and test cases. Our statistical significance tests reveal that NetML improves upon state-of-the-art bug localization approaches by a substantial margin.
\end{enumerate}


\subsection{Paper Organization}
\label{sec:paper_structs}

The remainder of this paper is organized as follows. In Section~\ref{sec.prelim}, we present background information on IR-based and spectrum-based bug localization approaches. Section~\ref{sec:approach} elaborates the proposed NetML in greater details. In Section~\ref{sec.exp}, we present our dataset, evaluation metrics, and experiment results. \recheck{Section~\ref{sec:qualitative} then provides a qualitative study of the NetML results, followed by discussions on potential threats to the validity of our study in Section~\ref{sec:threats}.  Section~\ref{sec.related} provides an overview of key related works. We finally conclude this paper and discuss future works in Section~\ref{sec.conclusion}.}

\section{Background}
\label{sec.prelim}

In this section, we present some background material on IR-based and spectrum-based bug localization.



\subsection{IR-Based Bug Localization}
\label{sec.IR-based} 

IR-based bug localization techniques consider an input bug report (i.e., the text in the summary and description of the bug report as a query, and program elements in a code base as documents, and employ IR techniques to sort the program elements based on their relevance with the query. The intuition behind these techniques is that program elements sharing many common words with the input bug report are likely to be relevant to the bug. By using text retrieval models, IR-based bug localization computes the similarities between various program elements and the input bug report. Then, program elements are sorted in descending order of their textual similarities to the bug report, and sent to developers for manual inspection.

%
%
%
%



All IR-based bug localization techniques need to extract textual contents from source code files and preprocess textual contents (either from bug reports or source code files). First, comments and identifier names are extracted from source code files. These can be extracted by employing a simple parser. In this work, we use JDT~\cite{jdt_link} to recover the comments and identifier names from source code. Next, after the textual contents from source code and bug reports are obtained, we need to preprocess them. The purpose of text preprocessing is to standardize words in source code and bug reports. There are three main steps: text normalization, stopword removal, and stemming:

\begin{enumerate}
	\item Text normalization breaks an identifier into its constituent words (tokens), following camel casing convention. Following the work by Saha et al.~\cite{SahaLKP13}, we also keep the original identifier names.
	\item Stopword removal removes punctuation marks, special symbols, number literals, and common stopwords~\cite{stopword_link}. It also removes programming keywords such as $\mathit{if}$, $\mathit{for}$, $\mathit{while}$, etc., which usually appear too frequently to be useful to differentiate between documents.
	\item Stemming simplifies English words into their root forms. For example, ''processed``, ''processing``, and ''processes`` are all simplified to ''process``. This increases the chance of a query and a document to share some common words. We use the popular Porter Stemming algorithm~\cite{P80}.
\end{enumerate}

Numerous IR techniques have been employed for bug localization. We highlight a popular IR technique namely \emph{Vector Space Model} (VSM). In VSM, queries and documents are represented as vectors of weights, where each weight corresponds to a term. The value of each weight is usually the \textit{term frequency\textemdash inverse document frequency} (TF-IDF)~\cite{Ramos1999} of the corresponding word. Term frequency refers to the number of times a word appears in a document. Inverse document frequency refers to the number of documents in a corpus (i.e., a collection of documents) that contain the word. The higher the term frequency and inverse document frequency of a word, the more important the word would be. In this work, given a document $d$ and a corpus $C$, we compute the TF-IDF weight of a word $w$ as follows:
\begin{align}
weight(w,d) &= \text{TF-IDF}(w,d,C) \nonumber\\
            &= \log(f(w,d)+1) \times \log\frac{|C|}{|{d_i\in C : w \in d_i }|} \nonumber
\end{align}
where $f(w,d)$ is the number of times $w$ appears in $d$.


%
%

After computing a vector of weights for the query and each document in the corpus, we calculate the cosine similarity of the query and document vectors. The cosine similarity between query $q$ and document $d$ is given by:
\begin{align}
\label{eqn:cosine_sim}
\mathit{sim}(q,d) = \frac{\sum\limits_{w\in (q\bigcap d)} \mathit{weight}(w,q) \times \mathit{weight}(w,d)}{\sqrt{\sum\limits_{w\in q}\mathit{weight}(w,q)^{2}} \times \sqrt{\sum\limits_{w\in d}\mathit{weight}(w,d)^{2}}}
\end{align}
where $w\in (q\bigcap d)$ means word $w$ appears both in the query $q$ and document $d$. Also, $\mathit{weight}(w,q)$ refers to the weight of word $w$ in the query $q$'s vector. Similarly, $\mathit{weight}(w,d)$ refers to the weight of word $w$ in the document $d$'s vector.

%


\subsection{Spectrum-Based Bug Localization} 
\label{sec.spectrum-based}

Spectrum-based bug localization (SBBL)---also known as spectrum-based fault localization (SBFL)---takes as input a faulty program and two sets of test cases. One is a set of failed test cases, and the other one is a set of passed test cases. SBBL then instruments the target program, and records program spectra that are collected when the set of failed and passed test cases are run on the instrumented program. Each of the collected program spectrum contains information of program elements that are executed by a test case. Various tools can be used to collect program spectra as a set of test cases are run. In this work, we use Cobertura~\cite{cobertura_link}.

\begin{table}[!t]
	\caption{Raw Statistics for Program Element $e$.}
	\center
    \begin{tabular}{|r|c|c|}
    \hline
    ~                 & e is \textit{executed} & e is \textit{not executed} \\ 
    \hline
    \hline

    {unsuccessful} test & $n_f(e)$             & $n_f(\bar e)$                 \\ 
    {successful} test   & $n_s(e)$             &$n_s(\bar e)$              \\ \hline
    \end{tabular}
    \label{tab:sbfl_matrix}
\end{table}

Based on this spectra, SBBL typically computes some raw statistics for every program element. Tables \ref{tab:sbfl_matrix} and \ref{tab:sbfl_notations} summarize some raw statistics that can be computed for a program element $e$, given a program spectra $p$. These statistics are the counts of unsuccessful (i.e., failed), and successful (i.e., passed) test cases that execute or do not execute $e$. If a successful test case executes program element $e$, then we increase $n_s(e, p)$ by one unit. Similarly, if an unsuccessful test case executes program element $e$, then we increase $n_f(e, p)$ by one unit. SBBL uses these statistics to calculate the suspiciousness scores of each program element. The higher the suspiciousness score, the more likely the corresponding program element is the faulty element. After the suspiciousness scores of all program elements are computed, program elements are then sorted in descending order of their suspiciousness scores, and sent to developers for manual inspection.


Different SBBL techniques have used different formulas to calculate the suspiciousness scores. Among these techniques, Tarantula is a popular one~\cite{JH05}. Using the notation in Table \ref{tab:sbfl_notations}, the following is the formula that Tarantula uses to compute the suspiciousness score of program element $e$, given program spectra $p$:
\begin{align}
\label{eqn:tarantula}
Tarantula(e, p)=\frac{\frac{n_f(e, p)}{n_f(p)}}{\frac{n_f(e, p)}{n_f(p)}+\frac{n_s(e, p)}{n_s(p)}}	
\end{align}
The main idea of Tarantula is that program elements that are executed by failed test cases are more likely to be faulty than those that are not executed. Thus, Tarantula assigns a non-zero score to program element $e$ that has $n_f(e, p) > 0$. 

\begin{table}[tb]
	\centering
	\caption{Raw Statistic Description.}
    \begin{tabular}{|p{1.3cm}|p{5.5cm}|}
    \hline
    \textbf{Notation} & \textbf{Description} \\
    \hline \hline
  \multirow{2}{*}{$n_f(e, p)$}        & Number of unsuccessful test cases executing program element $e$ in program spectra $p$          \\ \hline
     \multirow{2}{*}{$n_f(\bar{e}, p)$}        & Number of unsuccessful test cases that do not execute program element $e$ in program spectra $p$           \\ \hline
    \multirow{2}{*} {$n_s(e, p)$}         & Number of successful test cases that execute program element $e$ in program spectra $p$           \\ \hline
     \multirow{2}{*}{$n_s(\bar{e}, p)$}        & Number   of successful test cases that do not execute program element $e$ in program spectra $p$    \\ \hline
      $n_f(p)$        & Total number of unsuccessful test cases           \\ \hline

    $n_s(p)$        & Total number of successful test cases          \\ \hline

    \end{tabular}
    \label{tab:sbfl_notations}
\end{table}
\section{Proposed Framework}
\label{sec:approach}

\begin{figure}[!t]
\centering
\includegraphics[width=0.5\textwidth]{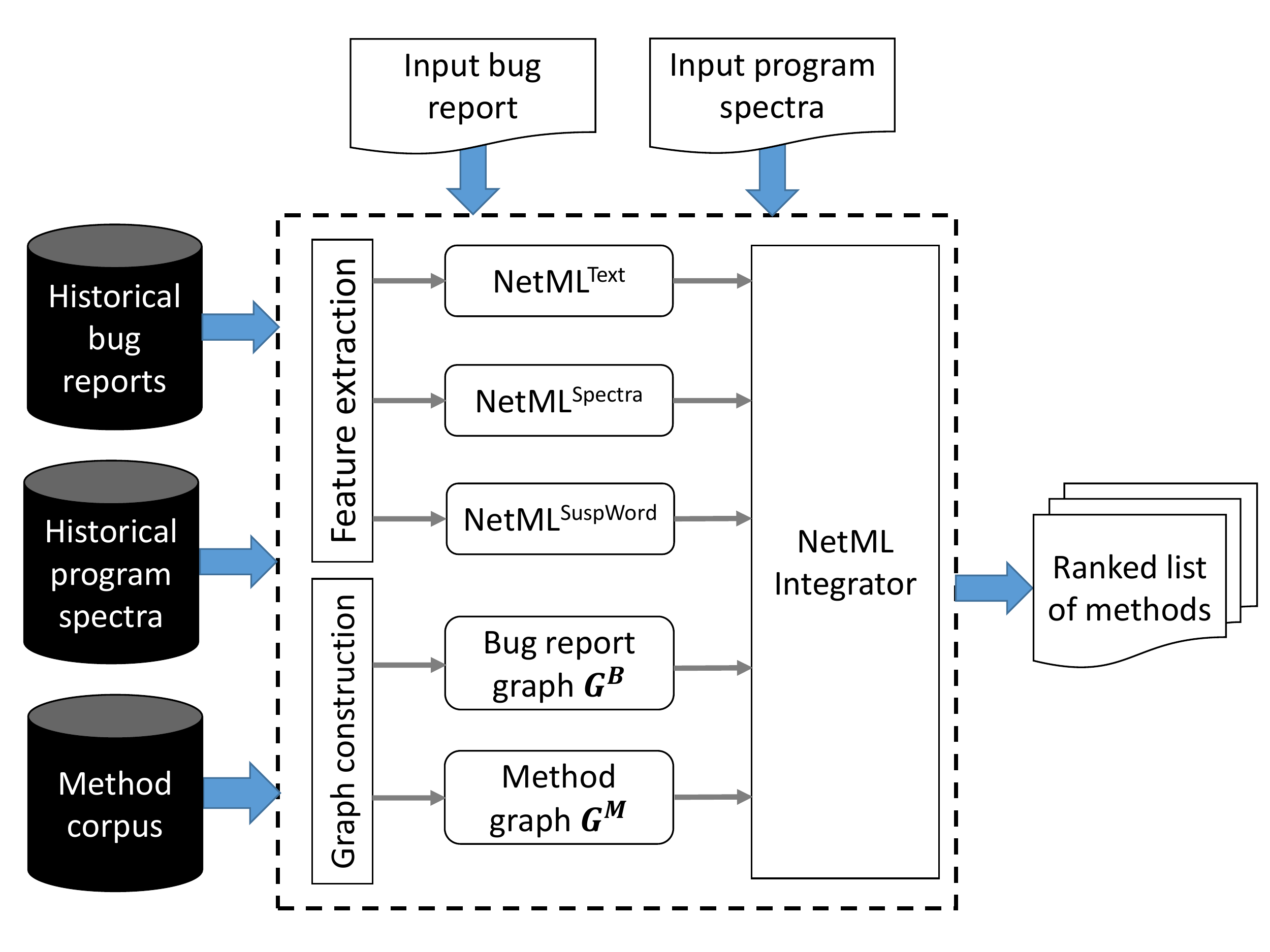}
\caption{The proposed NetML framework.}
\label{fig:framework}
\end{figure}

An overview of our NetML framework is given in Fig.~\ref{fig:framework} (enclosed in the dashed box). 
NetML takes as input a new bug report, the program spectra corresponding to it, and a method corpus. It also takes as input \emph{historical} bug reports that have been localized before. For each \textit{historical} bug report, we have its corresponding program spectra and ground truth labels. If a method contains a root cause of the bug, it is labeled as \textit{faulty}, otherwise it is labeled as \textit{non-faulty}. Given these inputs, NetML eventually produces a list of methods, ranked based on their likelihood to contain the root cause of the new bug report.


NetML has three main components, namely: \emph{feature extraction}, \emph{graph construction}, and \emph{integrator}. The feature extraction component serves to extract multi-modal input features that quantify different perspectives on the degree of relevancy between a bug report and a method. Note that this is in a similar spirit to~\cite{Le:2015:IRS:2786805.2786880}. Meanwhile, the graph construction component computes the similarity graphs among the bug reports ($\mathcal{G}_B$) and methods ($\mathcal{G}_M$).

Finally, the integrator component is the heart of NetML and constitutes the primary contribution of this work. It integrates both input features and similarity graph information in order to produce a ranked list of methods based on their relevancy score. In particular, the integrator performs adaptive learning that aims at jointly minimizing the bug localization errors and fostering clustering of the model parameters of similar bug reports and/or methods.

\recheck{In Sections \ref{subsec:feature} and \ref{subsec:graph}, we first elaborate the feature extraction and graph construction components respectively. We then describe the NetML integrator component in greater details in Sections \ref{sec.generalized_adaptive}--\ref{sec:learning}, including the formulation of our new integrator model as well as the corresponding objective function and adaptive learning procedure.} 


\subsection{Feature Extraction}
\label{subsec:feature}

The first component of the NetML framework is the feature extraction module, which generates features $\mathbf{X} = \{ x_{b,m.j} \}$ to be fed as inputs to the NetML integrator (see  Fig. \ref{fig:framework}). In line with our earlier AML work \cite{Le:2015:IRS:2786805.2786880}, for each bug report--method pair $(b,m)$, we compute a feature vector $\vec{x}_{b,m}$  that consists of three elements:
\begin{align}
\vec{x}_{b,m} = \left[ \text{NetML}^\text{Text}_{b,m}, \text{NetML}^\text{Spectra}_{b, m}, \text{NetML}^\text{SuspWord}_{b,m} \right]
\end{align}
The three features are elaborated in turn below.

$\text{NetML}^\text{Text}_{b,m}$ makes use of the TF-IDF method~\cite{Ramos1999} to estimate the similarity between methods and bug reports. In particular, given a method $m$ and a bug report $b$, $\text{NetML}^\text{Text}_{b,m}$ computes the cosine similarity between the TF-IDF representation of the bug report text and that of the method codes, which is akin to the IR-based bug localization method (cf. Section~\ref{sec.IR-based}). That is, $\text{NetML}^\text{Text}_{b,m}$ is given by:
\begin{align}
\text{NetML}^\text{Text}_{b,m} = sim(b,m)	
\end{align}
where $sim(b,m)$ is the cosine similarity as defined in (\ref{eqn:cosine_sim}).

$\text{NetML}^\text{Spectra}_{b, m}$ processes only the program spectra information using a spectrum-based bug localization technique described in Section~\ref{sec.spectrum-based}. Given a program spectra $p$ corresponding to bug report $b$ and a method $m$, $\text{NetML}^\text{Spectra}_{b, m}$ gives a score that quantifies how suspicious  $m$ is given $p$. By default, $\text{NetML}^\text{Spectra}_{b, m}$ uses the Tarantula method as described in Section \ref{sec.spectrum-based} (cf. equation (\ref{eqn:tarantula})):
\begin{align}
\text{NetML}^\text{Spectra}_{b, m} = Tarantula(m, p)
\end{align}

Finally, $\text{NetML}^\text{SuspWord}_{b,m}$ processes both bug reports and program spectra, and computes the suspiciousness scores of words to rank different methods. It breaks a method into its constituent words, computes the suspiciousness scores of these words, and then aggregates these scores back in order to arrive at the suspiciousness score of the method. Given a bug report $b$, a program spectra $p$, and a method $m$ in a corpus $C$, $\text{NetML}^\text{SuspWord}_{b,m}$ measures how suspicious $m$ is considering $b$ and $p$,  as follows: 
\begin{align}
&\text{NetML}^\text{SuspWord}_{b,m} = \text{NetML}^\text{Spectra}_{b, m} \times \\
&\frac{\sum\limits_{w\in b \cap m} \text{SSTFIDF}(w,p,b,C)\times \text{SSTFIDF}(w,p,m,C)}{\sqrt{\sum\limits_{w \in b} \text{SSTFIDF}(w,p,b,C)^2} \times \sqrt{\sum\limits_{w \in m} \text{SSTFIDF}(w,p,m,C)^2}}
\label{eq:sum_vsm_susp}
\end{align}
where 
$\text{SSTFIDF}(w,p,b,C)$ is the weight of a word $w$ in document (i.e., bug report or method) $d$ with corpus $C$ given program spectra $p$:
\begin{align}
\text{SSTFIDF}(w,p,d,C) = &\text{SS}_{\text{word}}(w,p)\times \ln(f(w,d)+1) \nonumber\\
&\times \ln\frac{|C|}{|{d_i\in C : w \in d_i }|}
\end{align}
where $\text{SS}_{\text{word}}(w,p)$ is the suspiciousness score of a word $w$:
\begin{align}
\text{SS}_{\text{word}}(w,p) &= \frac{\frac{|EF(w,p)|}{|p.FAIL|}}{\frac{|EF(w,p)|}{|p.FAIL|}+\frac{|ES(w,p)|}{|p.SUCCESS|}}
\label{eq:ss}
\end{align}
In the above equation, $EF(w,p)$ is the set of execution traces in $p.FAIL$ that contain a method in which the word $w$ appears, while $ES(w,p)$ is the set of execution traces in $p.SUCCESS$ that contain a method in which the word $w$ appears. Further details of all these components can be found in~\cite{Le:2015:IRS:2786805.2786880}.

\subsection{Graph Construction}
\label{subsec:graph}

The second component of the NetML framework is the graph construction module, which serves to compute the similarity graphs among bug reports and methods, to be used in the K-nearest neighbor retrieval as well as the network Lasso regularization. In this work, we define the bug report similarity graph $\mathcal{G}_B$ as comprising edge weights that reflect the textual similarity between two bug reports. For a pair of bug reports $b$ and $b'$, we define the edge weight $e_{b.b'}$ as follows:
\begin{align}
\label{eqn:edge_b}
e_{b,b'} = sim(b, b')
\end{align}
where $sim(b,b')$ is the cosine similarity between the TF-IDF weights of the textual descriptions of $b$ and $b'$, as per (\ref{eqn:cosine_sim}).

Similarly, the method similarity graph $\mathcal{G}_M$ comprises a set of edge weights $e_{m,m'}$ that reflect the textual similarity between two methods $m$ and $m'$. This is given by:
\begin{align}
\label{eqn:edge_m}
e_{m,m'} = sim(m, m')
\end{align}
where $sim(b,b')$ is the cosine similarity between the TF-IDF representations of the source codes of $m$ and $m'$.

\subsection{Integrator Model}
\label{sec.generalized_adaptive}


The new integrator model proposed in this work characterizes the relevancy of a method $m$ to a given bug report $b$ as an interaction between two types of model parameters, namely: \emph{ bug report parameters} $\vec{u}_b = [u_{b,1}, \ldots, u_{b,j}, \ldots, u_{b,J}]$ and \emph{ method parameters} $\vec{v}_m = [v_{m,1}, \ldots, v_{m,j}, \ldots, v_{m,J}]$, where $J$ is the total number of features. \recheck{Note that $J = 3$ in this case, i.e., $\text{NetML}^\text{Text}_{b,m}, \text{NetML}^\text{Spectra}_{b, m}, \text{NetML}^\text{SuspWord}_{b,m}$}. More specifically, the integrator model computes the relevancy score $\hat{f}_{b,m}$ as follows: 
\begin{align}
\label{eq:aml_plus}
\hat{f}_{b,m} = \hat{f}(\vec{x}_{b,m}, \vec{u}_b, \vec{v}_m) = \sum_{j = 1}^{J} (u_{b,j} + v_{m,j}) x_{b,m,j}
\end{align}
where $\vec{x}_{b,m} = [x_{b,m,1}, \ldots, x_{b,m,j}, \ldots, x_{b,m,J}]$ is the feature vector corresponding to a bug report--method pair $(b, m)$.

It is worth mentioning that the above model constitutes a generalization of the AML integrator model that we previously developed~\cite{Le:2015:IRS:2786805.2786880}. In AML, the final relevancy score is computed based solely on the  bug report parameters, and this set of parameters is shared by all methods for a given bug report. On the other hand, the NetML integrator model accounts for not only the  bug report parameters but also the  method parameters. The addition of the latter parameters provides a greater degree of freedom/flexibility in quantifying the contribution of different methods to the localization of a given bug report.

\subsection{Objective Function}
\label{subsec:objective}


Based on the above model formulation, we devise an objective function that guides the learning process of our integrator model. Specifically, we consider a joint optimization of bug localization quality and clustering of similar bug reports and methods, expressed by the loss function $\mathcal{L}$:
\begin{align}
\label{eq:NL_lossfunc}
\mathcal{L} &= \mathcal{L}_\text{Entropy} + \mathcal{L}_\text{Ridge} + \mathcal{L}_\text{NetLasso}
\end{align}
This consists of three components:
\begin{align}
\label{eq:entropy}
\mathcal{L}_\text{Entropy} = & -\sum_{b \in \mathcal{B}} \sum_{m \in \mathcal{M}} w_{b,m} \left[y_{b,m} \ln(\sigma(\hat{f}_{b,m})) \right. \nonumber\\
	& + \left. (1 - y_{b,m}) \ln(1- \sigma(\hat{f}_{b,m})) \right] \\
\label{eq:ridge}
\mathcal{L}_\text{Ridge} = &\frac{\alpha}{2} \sum_{j=1}^{J} \left[\sum_{b \in \mathcal{B}} u_{b,j}^2 + \sum_{m \in \mathcal{M}} v_{m,j}^2 \right] \\
\label{eq:netlasso}
\mathcal{L}_\text{NetLasso} = & \frac{\beta}{2} \sum_{j=1}^{J} \left[ \sum_{(b, b') \in \mathcal{G}^B}^{} e_{b, b'} (u_{b,j} - u_{b', j})^2 \right. \nonumber\\ 
	& + \left. \sum_{(m, m') \in \mathcal{G}^M}^{} e_{m, m'} (u_{m,j} - u_{m', j})^2 \right]
\end{align}
where $\mathcal{B}$ and $\mathcal{M}$ are the sets of bug reports and methods respectively, $y_{b,m}$ is a binary label that indicates whether method $m$ is relevant to bug report $b$ ($y_{b,m} = 1$) or not ($y_{b,m} = 0$), and $\sigma(\hat{f}_{b,m}) = \frac{1}{1 + \exp(-\hat{f}_{b,m})}$ is the logistic function \cite{Collins:2002:LRA:599615.599689}. Also, $w_{b,m}$ denotes the instance weight of a bug report--method pair $(b,m)$, while $e_{b,b'}$ and $e_{m,m'}$ are the edge weights reflecting the degree of similarity between two bug reports $b$ and $b'$, and two methods $m$ and $m'$, respectively. Finally, $\alpha > 0$ and $\beta > 0$ are the user-defined parameters that control the strength of the ridge and network Lasso regularization, respectively.

Note that $\mathcal{L}_\text{Entropy}$ refers to the so-called cross-entropy loss \cite{Murphy:2012:MLP:2380985}, which provides an error measure of the bug localization process. Here $\mathcal{L}_\text{Entropy}$ can be interpreted as the discrepancy between the probability distribution of the predictive model $\hat{f}_{b,m}$ and that of the true label $y_{b,m}$ \cite{Murphy:2012:MLP:2380985}. We also introduce the instance weight\footnote{An instance refers to a specific bug report--method pair $(b,m)$} $w_{b,m}$ in (\ref{eq:entropy}) to cater for the extremely \emph{skewed} distribution of the relevant vs. irrelevant methods for a given bug report, which is a major challenge in bug localization process. That is, the number of relevant (faulty) methods is much smaller than that of irrelevant (non-faulty) ones. To address this, we configure $w_{b,m}$ in such a way that imposes a greater penalty for relevant instances being incorrectly predicted/classified than that for irrelevant ones. Specifically, we set $w_{b,m}$ as:
\begin{align}
w_{b,m} =
\begin{cases}
    \frac{1}{N_\text{faulty}},     & \text{if } y_{b,m} = 1\\
    \frac{1}{N - N_\text{non-faulty}}, & \text{if } y_{b,m} = 0
\end{cases}
\end{align}
where $N$ is the total number of instances observed in the historical data, and $N_\text{faulty}$ is the number of faulty instances.

Meanwhile, the ridge regularization $\mathcal{L}_\text{Ridge}$ serves to penalize large values of the model parameters \cite{Murphy:2012:MLP:2380985}, which in turn helps mitigate the risk of data overfitting. From a probabilistic perspective, this corresponds to the Gaussian prior distribution for the model parameters $u_{b,j}$ and $v_{m,j}$, with zero mean and inverse variance of $\alpha$~\cite{Le:2015:IRS:2786805.2786880}. Finally, $\mathcal{L}_\text{NetLasso}$ refers to the network Lasso regularization \cite{Hallac:2015:NLC:2783258.2783313}, which enforces clustering of the model parameters of bug reports and methods. The intuition is straightforward---the more similar two bug reports or two methods are (as quantified by $e_{b,b'}$ and $e_{m,m'}$), the closer their model parameters $\vec{u}_b$ and $\vec{v}_m$ should be. This combination of $\mathcal{L}_\text{Entropy}$, $\mathcal{L}_\text{Ridge}$ and $\mathcal{L}_\text{NetLasso}$ facilitates a robust model that can simultaneously optimize the bug localization quality and cluster the model parameters of similar bug reports and methods.

Next, in order to minimize the joint loss $\mathcal{L}$, we employ a Newton method \cite{doi:10.1137/1.9780898718898} that is derived from a second-order Taylor series expansion of the loss function $\mathcal{L}$:
\begin{align}
\label{eq:newton}
	\mathcal{L}(\theta) = \mathcal{L}(\theta_0) + \triangledown \mathcal{L}(\theta_0) (\theta - \theta_0) + \frac{\triangledown^2 \mathcal{L}(\theta_0)}{2} (\theta - \theta_0)^2
\end{align}
The minima of $\mathcal{L}$ can be obtained by taking the partial derivative of $\mathcal{L}(\theta)$ and equating it to zero:
\begin{align}
0 &= \triangledown \mathcal{L}(\theta_0) + \triangledown^2 \mathcal{L}(\theta_0) (\theta - \theta_0) \nonumber\\
\theta &= \theta_0 - \frac{\triangledown \mathcal{L}(\theta_0)}{\triangledown^2 \mathcal{L}(\theta_0)}
\end{align}
If we take $\theta_0$ as the old estimate of $u_{b,j}$ or $v_{m,j}$, this leads to the following update formulae:
\begin{align}
\label{eqn:update_u}
u_{b,j} &\leftarrow u_{b,j} - \frac{ \triangledown \mathcal{L}(u_{b,j}) }{ \triangledown^2 \mathcal{L}(u_{b,j}) } \\
\label{eqn:update_v}
v_{m,j} &\leftarrow v_{m,j} - \frac{ \triangledown \mathcal{L}(v_{m,j}) }{ \triangledown^2 \mathcal{L}(v_{m,j}) }
\end{align}

In turn, we need to compute the first and second derivatives of each model parameter $u_{b,j}$ and $v_{m,j}$. For the  bug report parameter $u_{b,j}$, the first and second derivatives are respectively given by:
\begin{align}
\label{eqn:grad_u}
\triangledown \mathcal{L}(u_{b,j}) &= \sum_{m \in \mathcal{M}} \left[ w_{b,m} (\sigma(\hat{f}_{b,m}) - y_{b,m}) x_{b,m,j} \right] \nonumber \\
&+ \alpha u_{b,j} + \beta \sum_{b'} \left[ e_{b,b'} \left( u_{b,j} - u_{b',j} \right) \right] \\
\label{eqn:hess_u}
\triangledown^2 \mathcal{L}(u_{b,j}) &= \sum_{m \in \mathcal{M}} \left[ w_{b,m} \sigma(\hat{f}_{b,m}) ( 1 - \sigma(\hat{f}_{b,m}) ) x_{b,m,j}^2 \right] \nonumber \\
&+ \alpha + \beta \sum_{b'} e_{b,b'}
\end{align}
Similarly, we can compute the first and second derivatives w.r.t each  method parameter $v_{m,j}$ as:
\begin{align}
\label{eqn:grad_v}
\triangledown \mathcal{L}(v_{m,j}) &= \sum_{b \in \mathcal{B}} \left[ w_{b,m} (\sigma(\hat{f}_{b,m}) - y_{b,m}) x_{b,m,j} \right] \nonumber \\
& + \alpha v_{m,j} + \beta \sum_{m'} \left[ e_{m,m'} \left( v_{m,j} - v_{m',j} \right) \right] \\
\label{eqn:hess_v}
\triangledown^2 \mathcal{L}(v_{m,j}) &= \sum_{b \in \mathcal{B}} \left[ w_{b,m} \sigma(\hat{f}_{b,m}) ( 1 - \sigma(\hat{f}_{b,m})) x_{b,m,j}^2 \right] \nonumber \\
& + \alpha + \beta \sum_{m'} e_{m,m'}
\end{align}

\begin{algorithm*}[!ht]
	\begin{algorithmic}[1]
		\Require
		    \Statex Set of $K$ relevant historical bug reports $\mathcal{B}_K$ (i.e.,  $|\mathcal{B}_K| = K$)
			\Statex Set of all methods $\mathcal{M}$, where $|\mathcal{M}| = M$
		    \Statex New bug report query $b^*$ along with its features $\mathbf{X}_{b^*} = \{ x_{b^*,m,j} \} \in \mathbb{R}^{1 \times M \times J}$
			\Statex Historical features $\mathbf{X} = \{ x_{b,m,j} \} \in \mathbb{R}^{K \times M \times J}$
			\Statex Historical labels $\mathbf{Y} = \{ y_{b,m} \} \in \mathbb{R}^{K \times M}$
			\Statex Bug report similarity graph $\mathcal{G}_B$, represented by the adjacency matrix $\mathbf{E}_B = \{ e_{b,b'} \}$
			\Statex Method similarity graph $\mathcal{G}_M$, represented by the adjacency matrix $\mathbf{E}_M = \{ e_{m,m'} \}$
		\Ensure 
		    \Statex Relevancy scores $\hat{f}_{b^*,m} \in \mathbb{R}^{1 \times M}$ of the new bug report $b^*$ to all methods $m$
			\Statex Bug report parameters $\mathbf{U} = \{ u_{b,j} \} \in \mathbb{R}^{(K + 1) \times J}$ 
			\Statex Method parameters $\mathbf{V} = \{ v_{m,j} \} \in \mathbb{R}^{M \times J}$ 
		\Statex \hrulefill
		\State Compute the union set of bug reports $\mathcal{B} \leftarrow \mathcal{B}_K \cup \{ b^* \}$
		\State Initialize all model parameters $u_{b,j} \leftarrow 0$ and $v_{m,j} \leftarrow 0$, $\forall b \in \mathcal{B}, m \in \mathcal{M}, j \in \{ 1,\ldots,J \}$
		\State Precompute all constant terms $q_b \leftarrow \sum_{b'} e_{b,b'}$ and $q_m \leftarrow \sum_{m'} e_{m,m'}$, $\forall b \in \mathcal{B}, m \in \mathcal{M}$
		\State Compute the bug probabilities $\sigma(\hat{f}_{b,m})$ for all $(b,m)$ pairs via equation (\ref{eq:aml_plus})
		\State $\mathcal{L}_\text{curr} \leftarrow -\sum_{b} \sum_{m} w_{b,m} \big[ y_{b,m} \ln \big( \sigma(\hat{f}_{b,m}) \big) + \big( 1 - y_{b,m} \big) \ln \big(1 - \sigma(\hat{f}_{b,m}) \big) \big]$
		\Repeat 
		\State $\mathcal{L}_\text{prev} \leftarrow \mathcal{L}_\text{curr}$
		\For {each $j \in \{ 1,\ldots,J \}$}
		\State \textbf{/* Update all bug report parameters $u_{b,j}$ */}
		\For {each $b \in \mathcal{B}$}
		\State $p_{b} \leftarrow \sum_{b'} e_{b,b'} u_{b',j}$
		\EndFor
		\For {each $b \in \mathcal{B}$}
		\State $u_\text{numer} \leftarrow \sum_m \big[ w_{b,m} (\sigma(\hat{f}_{b,m}) - y_{b,m}) x_{b,m,j} \big] + \beta \big[ u_{b,j} q_b - p_{b} \big] + \alpha u_{b,j}$
		\State $u_\text{denom} \leftarrow \sum_m \big[ w_{b,m} \sigma(\hat{f}_{b,m}) (1 - \sigma(\hat{f}_{b,m})) x_{b,m,j}^2 \big] + \beta q_b + \alpha$
		\State $u_{b,j} \leftarrow u_{b,j} - \eta \left( \frac{u_\text{numer}}{u_\text{denom}} \right)$
		\EndFor
		\State \textbf{/* Update all method parameters $v_{m,j}$ */}
		\For {each $m \in \mathcal{M}$}
		\State $p_{m} \leftarrow \sum_{m'} e_{m,m'} v_{m',j}$
		\EndFor
		\For {each $m \in \mathcal{M}$}
		\State $v_\text{numer} \leftarrow \sum_b \big[ w_{b,m} (\sigma(\hat{f}_{b,m}) - y_{b,m}) x_{b,m,j} \big] + \beta \big[ v_{m,j} q_m - p_{m} \big] + \alpha v_{m,j}$
		\State $v_\text{denom} \leftarrow \sum_b \big[ w_{b,m} \sigma(\hat{f}_{b,m}) (1 - \sigma(\hat{f}_{b,m})) x_{b,m,j}^2 \big] + \beta q_m + \alpha$
		\State $v_{m,j} \leftarrow v_{m,j} - \eta \left( \frac{v_\text{numer}}{v_\text{denom}} \right)$
		\EndFor
		\EndFor
		\State Compute the updated bug probabilities $\sigma(\hat{f}_{b,m})$ via equation (\ref{eq:aml_plus})
		\State $\mathcal{L}_\text{curr} \leftarrow -\sum_{b} \sum_{m} w_{b,m} \big[ y_{b,m} \ln \big( \sigma(\hat{f}_{b,m}) \big) + \big( 1 - y_{b,m} \big) \ln \big(1 - \sigma(\hat{f}_{b,m}) \big) \big]$
		\State $\eta \leftarrow
		\begin{cases}
		\frac{\eta}{2},   & \text{if } \mathcal{L}_\text{curr} > \mathcal{L}_\text{prev}\\
		\min(1, 2 \eta),  & \text{otherwise}
		\end{cases}$
		\Until $T_{max}$ iterations
		\State Compute the relevancy scores $\hat{f}_{b^*,m}$ using equation (\ref{eq:aml_plus})
	\end{algorithmic}
	\caption{Adaptive learning of the NetML integrator}
	\label{alg:network_lasso}
\end{algorithm*}



Finally, the update formula for $u_{b,j}$ can be obtained by substituting equation (\ref{eqn:grad_u}) and (\ref{eqn:hess_u}) into equation (\ref{eqn:update_u}). Likewise, we can substitute (\ref{eqn:grad_v}) and (\ref{eqn:hess_v}) into (\ref{eqn:update_v}) to arrive at the update formula for $v_{m,j}$. To learn the model parameters, we use a \emph{Newton method} that updates the parameters on a per-feature $j$ basis. This will be elaborated in Section \ref{sec:learning}.

\subsection{Adaptive Learning}
\label{sec:learning}

Algorithm \ref{alg:network_lasso} summarizes the adaptive learning procedure of the NetML integrator for computing the relevancy scores of a new bug report (i.e., a new query) to different methods (i.e., documents). Given a new bug report $b^*$, the set of $K$ relevant bug reports $\mathcal{B}_K$ in the historical data,  the set of all methods $\mathcal{M}$, and the similarity graphs $\mathcal{G}_B$ and $\mathcal{G}_M$, the learning procedure appends $b^*$ into $\mathcal{B}_K$ and then updates the model parameters on a per-feature basis. That is, for each feature $j$, it performs Newton updates on the  bug report parameters $u_{b,j}$ (steps 14--16) and  method parameters $v_{m,j}$ (steps 23--25), in accordance with equations (\ref{eqn:update_u}) and (\ref{eqn:update_v}) respectively. \recheck{The key idea here is to alternatingly update the parameter for one feature while keeping the parameters of the remaining features fixed.} The procedure is repeated until a maximum iteration $T_{max}$ is reached. Afterwards, the final prediction score $\hat{f}_{b^*,m}$ of the new bug report $b^*$ for each method $m$ is computed via equation (\ref{eq:aml_plus}).

Note that the selection of relevant bug report set $B_K$ is based on the $K$-nearest neighbor retrieval from the bug report similarity graph $\mathcal{G}_B$, as follows:
\begin{align}
\mathcal{B}_K = \text{Top}_K( \{ e_{b^*, b} \mid \forall b \neq b^* \} )
\end{align}
where $\text{Top}_K$ is a function that returns bug reports with the highest similarity $e_{b^*, b}$ to the query bug report $b^*$. The calculation of the similarity graphs is based on the VSM model and will be further described in Section \ref{subsec:graph}.

It is also worth mentioning that the magnitude of the Newton update is downscaled by an adaptive learning rate $\eta$ (where $0 < \eta \leq 1$). We introduce this scaling factor as a way to address the problem of \emph{overshooting} in Newton method \cite{Bank1981}, whereby the update $\frac{ \triangledown \mathcal{L}(u_{b,j}) }{ \triangledown^2 \mathcal{L}(u_{b,j}) }$ or $\frac{ \triangledown \mathcal{L}(v_{m,j}) }{ \triangledown^2 \mathcal{L}(v_{m,j}) }$ is overestimated---possibly by many orders of magnitude. This may lead to oscillations and sometimes divergence in the loss function. To alleviate this issue, we compare the loss function $\mathcal{L}$ before and after a Newton iteration (step 30), and then adjust $\eta$ accordingly depending on whether $\mathcal{L}$ increases or not. If it increases, then we reduce $\eta$ by half in order to dampen the update magnitude; otherwise, the value of $\eta$ gets doubled, up to a maximum limit of $1$.

For computational efficiency, we precompute the constant terms $q_b = \sum_{b'} e_{b,b'}$ and $q_m = \sum_{m'} e_{m,m'}$ before the Newton iterations begins. Additionally, during each Newton iteration, we have separate loops to compute the terms $\sum_{b'} e_{b,b'} u_{b',j}$ (step 11) and $\sum_{m'} e_{m,m'} v_{m',j}$ (step 20) for each feature $j$, prior to updating $u_{b,j}$ and $v_{m,j}$. The purpose is to make sure that, during the parameter updates (steps 14 and 23), the computation of $\sum_{b'} e_{b,b'} u_{b',j}$ in equation (\ref{eqn:update_u}) and $\sum_{m'} e_{m,m'} v_{m',j}$ in equation (\ref{eqn:update_v}) is based on the old parameter values from the previous iteration, and not affected by the ordering of $b$ or $m$ in the update loops.



We additionally highlight that the loss function $\mathcal{L}$ is \emph{strictly convex}. This provides a nice theoretical guarantee that there is only one unique minimum in the loss function surface, and this minimum is globally optimal \cite{Renegar:2001:MVI:502968}. The convexity trait can be proven by looking at the curvatures (i.e., second derivatives) with respect to the  bug report and method parameters, as per equations (\ref{eqn:hess_u}) and (\ref{eqn:hess_v}) respectively. Clearly, since $0 \leq \sigma(\hat{f}_{b,m}) \leq 1$, $w_{b,m}$, $x_{b,m,j}^2$, $e_{b,b'}$ and $e_{m,m'}$ are non-negative, while $\alpha$ and $\beta$ are positive, the curvatures will be positive. The positive curvatures correspond to the so-called \emph{positive definite} Hessian matrix---a well-known property of a strictly convex function \cite{Renegar:2001:MVI:502968}.

\section{Experiments }
\label{sec.exp}

In this section, we first describe the datasets and evaluation settings used in our experiments. We then present a list of research questions we want to address, and accordingly elaborate our experiment results. 


\subsection{Dataset}\label{sec:dataset}


To evaluate our approach, we use a dataset of 355 bugs from seven popular software projects. The seven projects are Ant~\cite{ant_link}, AspectJ~\cite{aspectj_link}, Lang~\cite{lang_link}, Lucene~\cite{lucene_link}, Math~\cite{math_link}, Rhino~\cite{rhino_link}, and Time~\cite{time_link}. All seven projects are medium-large scale and implemented in Java. Ant, AspectJ, and Lucene contain more than 300 KLOC. Math, Rhino, and Time contains almost 100 KLOC, while Lang only contains more than 50 KLOC. Ant, Lang, Lucene, and Math projects use Jira as the issue tracking system, from which we retrieve their bug reports. Bissyande et al. found that in Jira bugs are generally well linked to commits that fix them~\cite{BissyandeTWLJR13}. AspectJ and Rhino uses Bugzilla whereas Time uses Github as the issues tracking system, from which we collect their bug reports. Table~\ref{tab:dataset} presents an overview of the seven projects considered in our study.

The 116 bugs from Ant, Lucene, and Rhino were collected by ourselves, following the procedure used in \cite{DallmeierZ07}. For each bug, we collected the pre-fix version, post-fix version, a set of successful test cases, and at least one failing test case. A failing test case is often included as an attachment to a bug report or committed along with the fix in the post-fix version. When a developer receives a bug report, he/she first needs to replicate the error described in the report~\cite{worksforme}. In this process, he is creating a failing test case. Unfortunately, not all test cases are documented and saved in the version control systems. The 41 AspectJ bugs are from the iBugs dataset which was collected by Dallmeier and Zimmermann~\cite{DallmeierZ07}. Each bug in the iBugs dataset comes with the code before the fix (pre-fix version), the code after the fix (post-fix version), and a set of test cases. The iBugs dataset contains more than 41 AspectJ bugs, but not all of them come with failing test cases. Test cases provided in the iBugs dataset are obtained from the various versions of the regression test suite that comes with AspectJ. We collected the remaining 198 bugs from Lang, Math, and Time from Defects4J benchmark~\cite{Just:2014:DDE:2610384.2628055}, a database of real, isolated, reproducible software faults from real-world open-source Java projects. The three projects include a large number of test cases, and there exists at least one failing test case per bug. Defects4J also contains two other projects, namely JFreechart and Closure-Compiler. We omit these projects since we are unable to fully collect all their bug reports. 

\begin{table}[t!]
	\centering 
	\caption{Summary of the datasets used in this work. \recheckagain{We use the short names of projects for brevity; ``Ant'' stands for ``Apache-Ant'', ``Lang'' stands for ``Apache-Commons-Lang'', ``Math'' stands for ``Apache-Commons-Math'', and ``Time'' stands for ``Joda-Time''}.}
    \begin{tabular}{|l|c|c|r|}
    \hline
   \multirow{2}{*}{\textbf{Project}} &\multirow{2}{*}{\textbf{\#Bugs}}&{\multirow{2}{*}{\textbf{Time Period}}}& \textbf{Average}   \\
    ~       &             &                                   &\textbf{\# Methods}   \\ \hline
	\hline
	Ant       & 53                                          & 12/2001 -- 09/2013 & 9,624.66   \\ 
    AspectJ       & 41                                              & 03/2005 -- 02/2007                                   & 14,218.39   \\ 
     
Lang       &     65                                          & 	10/2002 -- 	04/2016 & 2,151.1  \\
    Lucene       & 37                                              & 06/2006 -- 01/2011                   & 10,220.14   \\
    Math       & 106                                              &  12/2004 --  03/2016& 4,792.3   \\ 
    Rhino       & 26                                                 &  12/2007 -- 12/2011                                   & 4,839.58   \\
    Time       & 27                                                 &                      05/2004 -- 03/2017            & 4,083.5
    \\ \hline
    \end{tabular}
    \label{tab:dataset}
\end{table}

\subsection{Evaluation Metrics and Settings}
\label{sec:metrics}

To assess the effectiveness of a bug localization method, we employ two key metrics, namely: Top N and mean average precision (MAP). They are respectively described below:

\begin{itemize}
	\item  \textbf{Top N}: Given a bug report, if one of its corresponding faulty methods is in the top-N results, we consider that the bug is successfully localized. The Top N score of a bug localization method is the number of bugs it can successfully localize~\cite{Zhou:2012:BFM:2337223.2337226,SahaLKP13}.
	\item  \textbf{Mean Average Precision (MAP)}: MAP is an IR metric to evaluate ranking approaches~\cite{Manning2008}, and is computed by taking the mean of the {\em average precision} scores across all bug reports. The average precision of a single bug report is computed as:
	\[
	AP = \frac{\sum_{k=1}^{M}P(k) \times pos(k)}{\sum_{k=1}^{M} pos(k)}
	\]
	where $k$ is a rank in the returned ranked methods, $M$ is the number of ranked methods, and $pos(k)$ indicates whether the $k^{th}$ method is faulty or not. Here $P(k)$ is the precision at a given top $k$ methods, which is computed as follows:
	\[
	P(k)=\frac{\#faulty\ methods\ in\ the\ top\ k}{k}.
	\]
Note that the MAP scores of existing bug localization methods are typically low~\cite{Rao:2011:RSL:1985441.1985451,Sisman:2012:IVH:2664446.2664454,Zhou:2012:BFM:2337223.2337226,SahaLKP13}.
\end{itemize}



Our evaluation procedure is based on 10-fold \emph{cross validation} (CV). That is, for each project, we divide the bug reports into ten (mutually exclusive) sets. Then, for each fold, we take 1 set as new bug report queries (i.e., testing set) and treat the remaining 9 sets as historical bug reports (i.e., training set). We repeat this 10 times, and then collate the results to get the aggregated Top N and MAP scores.
	
In all our experiments, the hyper-parameters of the NetML method were configured as follows. Firstly, the regularization parameters $\alpha$ and $\beta$ were chosen by performing 10 fold cross validation on the training set. Next, the maximum number of iterations $T_{max}$ was fixed to $30$. We use $K=10$ as default value for the number of nearest neighbors. Note that the NetML parameters $K$ and $T_{max}$ follow the settings used in AML~\cite{Le:2015:IRS:2786805.2786880}, so as to ensure fair comparisons. All experiments were conducted on an Intel(R) Xeon 2.9GHz server running a Linux operating system.

In order to assess whether NetML substantially outperforms other bug localization methods, we apply \emph{Wilcoxon signed-rank test}~\cite{Wilcoxon1945}; it is a non-parametric statistical significance test for comparing two related or matched samples, whereby the population cannot be assumed to be normally distributed. The Wilcoxon test was applied to two types of metric (i.e., Top N and MAP). For every evaluation metric, we collated the 10-fold results of a bug localization technique across the four software projects (i.e., AspectJ, Ant, Lucene, and Rhino) and then performed the Wilcoxon test to compare the collated results of different techniques. For this test, our null hypothesis is that NetML performs \emph{worse} than or \emph{equal} to the other method, and so we used one-sided/tail $p$-value to validate this hypothesis. Moreover, we also apply the Benjamini-Hochberg (BH)~\cite{Benjamini1995thecontrol} procedure to control the effect of multiple comparisons. If the $p$-value is sufficiently small (say, below a significance level of $0.05$), we can confidently reject the null hypothesis and conclude that NetML is significantly better than the other method.

\subsection{Research Questions}
\label{sec:rqs}

Our empirical study seeks to answer several research questions (RQ), as described in the following subsections.

\subsubsection{RQ1: How Effective is NetML Compared to Other State-of-the-Art Techniques?}


We compare out NetML approach with its predecessor, i.e., AML~\cite{Le:2015:IRS:2786805.2786880}, and several other state-of-the-art techniques. Previously, Le et al. proposed Savant~\cite{B.Le:2016:LBF:2931037.2931049}, a state-of-the-art bug localization approach that employs a learning-to-rank~\cite{svmrank} strategy, using likely invariant \textit{diffs} and suspiciousness scores as features. Ochiai~\cite{Abreu:2007:ASF:1308173.1308264} and Dstar~\cite{DBLP:journals/tr/WongDGL14} are well-known statistical formulas to detect suspicious locations for bug localization without requiring any prior information on program structure or semantics. PROMESIR~\cite{PoshyvanykGMAR07}, SITIR~\cite{LiuMPR07}, and several variants developed by Dit et al.~\cite{DitRP13} were state-of-the-art multi-modal feature location techniques. Among the variants proposed by Dit et al.~\cite{DitRP13}, the best performing ones were $IR_{LSI}Dyn_{bin}WM_{HITS}(h,bin)^{bottom}$ and  $IR_{LSI}Dyn_{bin}WM_{HITS}(h,freq)^{bottom}$. In this paper, we refer to these variants as DIT$^\text{A}$ and DIT$^\text{B}$ respectively. Dit et al. had shown that these two variants outperform SITIR, and so we exclude SITIR from our study. We also compare NetML with a state-of-the-art IR-based bug localization method named LR~\cite{YeBL14}, and a state-of-the-art spectrum-based bug localization method named MULTRIC~\cite{XuanM14}. Note that, unlike PROMESIR, DIT$^\text{A}$, DIT$^\text{B}$, and MULTRIC which locate buggy methods, LR locates buggy files. Thus, we convert the list of files that LR produces into a list of methods by using two heuristics: (1) to return methods in a file in the same order that they appear in the file; and (2) to return methods based on their similarity to the input bug report as computed using a VSM model. We refer to the two variants of LR as LR$^{A}$ and LR$^{B}$ respectively.


For all the above-mentioned techniques, we used the same parameters and settings as described in the respective papers, with the following exceptions that we justify. For DIT$^\text{A}$ and DIT$^\text{B}$, the threshold used to filter methods using HITS was decided ``such that at least one gold set method remained in the results for 66\% of the [bugs]''~\cite{DitRP13}. In this paper, since we used 10-fold CV, rather than using 66\% of all bugs, we used all bugs in the training data (i.e., 90\% of all bugs) to tune the threshold. For PROMESIR, we also used 10-fold CV and applied a brute force approach to tune PROMESIR's component weights using a step of 0.05.

\subsubsection{RQ2: Do Feature Components of NetML Contribute toward Its Overall Performance?}


To answer this question, we conducted an ablation test by dropping one feature component (i.e., $\text{NetML}^\text{Text}$, $\text{NetML}^\text{SuspWord}$, or $\text{NetML}^\text{Spectra}$) one-at-a-time and evaluating the performance. In the process, we created three variants of NetML: $\text{All}^{-\text{Text}}$, $\text{All}^{-\text{SuspWord}}$ , and  $\text{All}^{-\text{Spectra}}$. That is, we excluded Text, SuspWord, and Spectra from all feature components, respectively (see also Fig. \ref{fig:framework}). We used the default value of $K=10$, and applied the NetML adaptive learning procedure (i.e., Algorithm \ref{alg:network_lasso}) to tune the model parameters of these variants. As our baseline, we performed the same ablation test to the feature components of the AML method (i.e., $\text{AML}^\text{Text}$, $\text{AML}^\text{SuspWord}$, or $\text{AML}^\text{Spectra}$).

\subsubsection{RQ3: How Effective is the NetML Integrator?}

Instead of using the NetML integrator component (see Section \ref{sec.generalized_adaptive}), one may consider using a standard machine learning algorithm, such as the learning-to-rank method, to combine the scores produced by the three feature components. Indeed, state-of-the-art IR-based and spectrum-based bug localization techniques such as LR and MULTRIC are based on the learning-to-rank method. As such, we conduct an experiment to compare our NetML integrator model with an off-the-shelf learning-to-rank model called SVM$^{rank}$~\cite{svmrank}, which was also used by LR~\cite{YeBL14}. To do so, we simply replace the NetML integrator model in Fig. \ref{fig:framework} with SVM$^{rank}$, and then compare the resulting performances. For completeness, we also compare our NetML integrator with the integrator model used by the AML algorithm.

\newcolumntype{C}[1]{>{\centering\arraybackslash}p{#1}}
\newcolumntype{L}[1]{>{\raggedright\arraybackslash}p{#1}}

%

\subsubsection{RQ4: What is the Effect of Varying the Number of Neighbors $K$ on the Performance of NetML?}

The most important parameter in our NetML approach is the number of nearest neighbors $K$ (while the regularization parameters $\alpha$ and $\beta$ were chosen via cross-validation---see Section \ref{sec:metrics}). By default, we set the number of neighbors to $K=10$, but the effect of varying this value is unclear. To answer this research question, we vary the value of $K$ and investigate its effects on the performance of NetML. In particular, we wish to investigate if the performance remains relatively stable with varying values of $K$. For each $K$ value, we also compare the performance of NetML against its predecessor (i.e., AML) using the same value.



%

\subsubsection{RQ5: How Effective is NetML in Cross-Project Bug Localization?}


\recheck{To evaluate the robustness of our approach, we also conducted an empirical study on cross-project bug localization. That is, we first use a source project as training data to build a bug localization model, and then employ the model to predict a method that likely contains a bug in a (different) target project~\cite{Wang:2016:ALS:2884781.2884804}. In this study, we compare NetML with its predecessor (i.e., AML)~\cite{Le:2015:IRS:2786805.2786880}, Savant~\cite{B.Le:2016:LBF:2931037.2931049}, Ochiai~\cite{Abreu:2007:ASF:1308173.1308264} and Dstar~\cite{DBLP:journals/tr/WongDGL14}. 
We use the same evaluation metrics as per Section~\ref{sec:metrics} to assess the effectiveness of the different techniques. 
To configure the hyper-parameters of NetML, we adopt the same parameter tuning procedure as described in Section~\ref{sec:metrics}. Meanwhile, the hyper-parameters of the remaining localization techniques follow the parameter settings stated in their respective papers. We also apply \textit{Wilcoxon signed-rank test} with the BH procedure to verify if NetML performs substantially better than the other techniques.}


\subsection{Results}
\label{sec:results}

This section presents our experiment results and discussion in relation to the research questions raised in Section \ref{sec:rqs}.

\subsubsection{RQ1: Comparisons of NetML with Other Techniques}
\label{sec:rq_benchmark}

\begin{table*}[t!]
	\centering
	\caption{\recheckagain{Top N (N $\in \{1, 5, 10\}$) results of NetML vs. AML, Savant, Ochiai, Dstar, and PROMESIR. The percentage in parentheses indicates the proportion of bug reports whose faulty methods are correctly localized}.}
	\begin{tabular}{|c|c|c|c|c|c|c|}
		\hline
		\textbf{Top N} & \textbf{NetML} & \textbf{AML} & \textbf{SAVANT} & \textbf{OCHIAI} & \textbf{DSTAR} & \textbf{PROMESIR} \\
		\hline
		\hline
		1     & \textbf{116 (32.68\%)} & 88 (24.79\%) & 67 (21.34\%) & 48 (13.52\%) & 43 (12.11\%) & 61 (17.18\%) \\
		5     & \textbf{219 (61.69\%)} & 179 (50.42\%) & 122 (38.85\%) & 94 (26.48\%) & 88 (24.79\%) & 139 (39.15\%) \\
		10    & \textbf{255 (71.83\%)} & 213 (60.00\%) & 152 (48.41\%) & 124 (34.93\%) & 106 (29.86\%) & 174 (49.01\%) \\
		\hline
	\end{tabular}%
	\label{tab:result_top_N}%
\end{table*}%

\begin{table*}[t!]
	\centering
	\caption{\recheckagain{Top N (N $\in \{1, 5, 10\}$) results of NetML vs. DIT$^\text{A}$, DIT$^\text{B}$, LR$^A$, LR$^B$, and MULTRIC. The percentage in parentheses indicates the proportion of bug reports whose faulty methods are correctly localized}.}
	\begin{tabular}{|c|c|c|c|c|c|c|}
		\hline
		\textbf{Top N} & \textbf{NetML} & \textbf{DIT$^\text{A}$} & \textbf{DIT$^\text{B}$} & \textbf{LR$^A$} & \textbf{LR$^B$} & \textbf{MULTRIC} \\
		\hline
		\hline
		1     & \textbf{116 (32.68\%)} & 41 (11.55\%) & 37 (10.42\%) & 12 (3.38\%) & 66 (18.59\%) & 68 (19.15\%) \\
		5     & \textbf{219 (61.69\%)} & 88 (24.79\%) & 78 (21.97\%) & 67 (18.87\%) & 137 (38.59\%) & 133 (37.46\%) \\
		10    & \textbf{255 (71.83\%)} & 117 (32.96\%) & 109 (30.7\%) & 116 (32.68\%) & 181 (50.99\%) & 162 (45.63\%) \\
		\hline
	\end{tabular}%
	\label{tab:result_top_N_}%
\end{table*}%

\begin{table*}[t!]
	\centering
	\caption{Mean Average Precision (MAP) results of different bug localization methods.}
	\begin{tabular}{|l|c|c|c|c|c|c|c|c|c|c|c|}
		\hline
		\multicolumn{1}{|l|}{\textbf{Project}} & \textbf{NetML} & \textbf{AML} & \textbf{Savant} & \textbf{Ochiai} & \textbf{Dstar} & \textbf{PROMESIR} & \textbf{DIT}$^\text{A}$  & \textbf{DIT}$^\text{B}$ & \textbf{LR}$^A$ & \textbf{LR}$^B$ & \textbf{MULTRIC} \\
		\hline\hline
		Ant   & 0.270  & 0.234 & 0.188 & 0.179 & 0.127 & 0.206 & 0.12  & 0.120  & 0.070  & 0.218 & 0.077 \\
		Aspectj & 0.219 & 0.187 & --    & 0.117 & 0.007 & 0.121 & 0.092 & 0.071 & 0.006 & 0.004 & 0.016 \\
		Lang  & 0.638 & 0.542 & 0.535 & 0.147 & 0.146 & 0.394 & 0.198 & 0.184 & 0.167 & 0.424 & 0.564 \\
		Lucene & 0.290  & 0.284 & 0.178 & 0.133 & 0.136 & 0.204 & 0.169 & 0.166 & 0.063 & 0.184 & 0.188 \\
		Math  & 0.358 & 0.255 & 0.261 & 0.14  & 0.139 & 0.271 & 0.179 & 0.176 & 0.165 & 0.303 & 0.391 \\
		Rhino & 0.302 & 0.243 & 0.243 & 0.137 & 0.127 & 0.203 & 0.092 & 0.09  & 0.034 & 0.103 & 0.172 \\
		Time  & 0.354 & 0.294 & 0.166 & 0.115 & 0.115 & 0.148 & 0.062 & 0.062 & 0.051 & 0.142 & 0.282 \\
		\hline
		\textbf{Overall} &\textbf{ 0.347} & 0.291 & 0.262 & 0.138 & 0.114 & 0.221 & 0.130  & 0.124 & 0.079 & 0.197 & 0.241 \\
		\hline
	\end{tabular}
	\label{tab:result_map}
\end{table*}

\begin{table*}[t!]
	\centering
	\caption{The $p$-values of the Wilcoxon test applying the BH procedure on various pairs of bug localization methods.}
	\begin{tabular}{|l|c|c|c|c|}
		\hline
		\textbf{Method Comparision} & \textbf{Top 1 } & \textbf{Top 5} & \textbf{Top 10} & \textbf{MAP} \\
		\hline\hline
		NetML vs. AML & $3 \times 10^{-7}$ (**) & $4 \times 10^{-5}$ (**) & 0.008 (**) & $5 \times 10^{-8}$ (**) \\
		NetML vs. Savant & $6 \times 10^{-8}$ (**) & $1 \times 10^{-5}$ (**) & $9 \times 10^{-4}$ (**) & $1 \times 10^{-8}$ (**) \\
		NetML vs. Ochiai & $2 \times 10^{-7}$ (**) & $4 \times 10^{-10}$ (**) & $6 \times 10^{-10}$ (**) & $6 \times 10^{-12}$ (**) \\
		NetML vs. Dstar & $1 \times 10^{-7}$ (**) & $8 \times 10^{-8}$ (**) & $1 \times 10^{-15}$ (**) & $1 \times 10^{-11}$ (**) \\
		NetML vs. PROMESIR & $8 \times 10^{-9}$ (**) & $1 \times 10^{-8}$ (**) & $5 \times 10^{-6}$ (**) & $4 \times 10^{-10}$ (**) \\
		NetML vs. DIT$^\text{A}$ & $4 \times 10^{-14}$ (**) & $2 \times 10^{-16}$ (**) & $3 \times 10^{-16}$ (**) & $8 \times 10^{-21}$ (**) \\
		NetML vs. DIT$^\text{B}$ & $4 \times 10^{-15}$ (**) & $8 \times 10^{-17}$ (**) & $1 \times 10^{-20}$ (**) & $3 \times 10^{-27}$ (**) \\
		NetML vs. LR$^A$ & $1 \times 10^{-18}$ (**) & $5 \times 10^{-22}$ (**) & $4 \times 10^{-20}$ (**)0 & $8 \times 10^{-22}$ (**) \\
		NetML vs. LR$^B$ & $4 \times 10^{-16}$ (**) & $2 \times 10^{-21}$ (**) & $2 \times 10^{-20}$ (**) & $1 \times 10^{-24}$ (**) \\
		NetML vs. MULTRIC & $3 \times 10^{-16}$ (**) & $1 \times 10^{-21}$ (**) & $1 \times 10^{-20}$ (**) & $1 \times 10^{-28}$ (**) \\
		\hline
		\multicolumn{5}{l}{(**): smaller than $0.01$}
	\end{tabular}%
	\label{tab:wilcoxon_methods}%
\end{table*}%

\rechecknewagain{Tables~\ref{tab:result_top_N} and~\ref{tab:result_top_N_} show the Top N results of NetML as well as the other baseline methods including AML. Out of the 355 bugs, NetML is able to successfully localize 116, 219, and 255 bugs when the developers inspect the Top 1, Top 5, and Top 10 methods respectively. This implies that NetML can successfully localize 31.82\%, 22.35\%, and 19.72\% more bugs than the best baseline (i.e., AML) by examining the Top 1, Top 5, and Top 10 methods respectively. For more details on the Top N results for each software project, please see Tables~\ref{tab:full_result_top_N} and~\ref{tab:full_result_top_N_} in the Appendix.
Note that we encountered java.lang.UnsupportedClassVersionError when running Savant for AspectJ bugs. These AspectJ bugs are from iBugs dataset~\cite{DallmeierZ07}. We have investigated and found that according to iBugs's documentation\footnote{ https://www.st.cs.uni-saarland.de/ibugs/}, the AspectJ's faulty versions work with Java Virtual Machine (JVM) version 1.4. However, Savant employs Daikon~\cite{Ernst:2007:DSD:1321774.1321800} which requires Java 7 or later\footnote{ http://plse.cs.washington.edu/daikon/download/doc/daikon.html}. Therefore, we exclude AspectJ's bugs from the experiments 
for Savant.}

Table~\ref{tab:result_map} shows the MAP score of NetML along with those of the state-of-the-art multi-modal localization methods. Averaging across the seven projects, NetML achieves an overall MAP score of 0.347, which outperforms all the other baselines. In particular, NetML improves the average MAP of AML, Savant, Ochiai, Dstar, PROMESIR, DIT$^\text{A}$, DIT$^\text{B}$, LR$^{A}$, LR$^{B}$, and MULTRIC by 19.24\%, 32.44\%, 151.45\%, 204.39\%, 57.01\%, 166.92\%, 62.15\%, 339.24\%, 76.14\% and 43.98\% respectively. Considering the individual projects, NetML remains the best performing approach in terms of MAP. That is, NetML achieves MAP scores of 0.270, 0.219, 0.638, 0.290, 0.358, 0.302, and 0.354 for the Ant, AspectJ, Lang, Lucene, Math, Rhino, and Time projects respectively. With respect to the best performing baseline (i.e., AML), these respectively constitute of 15.38\%, 17.11\%, 17.71\%, 2.11\%, 40.39\%, 24.28\%, and 20.41\% improvements.

We finally performed the Wilcoxon test to compare the Top N and MAP results of different techniques. As we are unable run Savant on AspectJ, we omit this project and run the Wilcoxon test on the results collated over the remaining six software projects for each metric (i.e., Top 1, Top 5, Top 10, and MAP). Table~\ref{tab:wilcoxon_methods} presents the $p$-values for the four metrics, evaluated at the significance levels of $0.05$ and $0.01$. The results show that NetML significantly outperforms AML on all the four metrics. Compared to the remaining techniques, NetML also performs significantly better in terms of Top 1, Top 5, Top 10 methods and MAP. Altogether, these results demonstrate the efficacy of the proposed NetML approach.

\subsubsection{RQ2: Contribution of Feature Components}
\label{sec:rq_feature}

\begin{table*}[t]
	\caption{Contributions of feature components in NetML and AML.  The percentage in parentheses indicates the propotion of bug reports whose faulty methods are correctly localized.}
	\centering
	%
	%
	%
	
	\begin{tabular}{|l|c|c|c|c|c|c|c|c|}
		\hline 
		\multirow{2}{*}{\textbf{Approach}} & \multicolumn{2}{c|}{\textbf{Top 1}} & \multicolumn{2}{c|}{\textbf{Top 5}} & \multicolumn{2}{c|}{\textbf{Top 10}} & \multicolumn{2}{c|}{\textbf{MAP}}\tabularnewline
		\cline{2-9} 
		& \textbf{NetML} & \textbf{AML} & \textbf{NetML} & \textbf{AML} & \textbf{NetML} & \textbf{AML} & \textbf{NetML} & \textbf{AML}\tabularnewline
		\hline 
		\hline
		All$^{-\text{Text}}$ & 68 (19.15\%)  & 61 (17.18\%) & 144 (40.56\%) & 130 (36.62\%) & 179 (50.42\%) & 165 (46.48\%) & 0.228 & 0.212\\
		All$^{-\text{Spectra}}$ & 56 (15.77\%) & 49 (13.80\%) & 128 (36.06\%) & 112 (31.65\%) & 172 (48.45\%) & 157 (44.23\%) & 0.215 & 0.210\\
		All$^{-\text{SuspWord}}$ & 74 (20.85\%) & 65 (18.31\%) & 156 (43.94\%) & 136 (38.31\%) & 196 (55.21\%) & 182 (51.27\%) & 0.211 & 0.229\\
		\hline
		All & \textbf{116} \textbf{(36.62\%)} & 88 (24.79\%) & \textbf{219} \textbf{(61.69\%)} & 179 (50.42\%) & \textbf{255} \textbf{(71.83\%)} & 213 (60.00\%) & \textbf{0.347} & 0.291\\
		\hline 
	\end{tabular}
	
	\label{tab:AML_variants}
\end{table*}

Table~\ref{tab:AML_variants} summarizes the results of our ablation test on both NetML and AML, each comparing the full model and three reduced variants (i.e., All$^{-\text{Text}}$, All$^{-\text{Spectra}}$ and All$^{-\text{SuspWord}}$). It is evident that, for both NetML and AML,  the full model always performs better than the reduced variants. This suggests that each feature component plays an important role, and omitting one of them may greatly affect the modelling performance. Among the three variants, it can be seen that {All}$^{-\text{SuspWord}}$ yields the smallest Top 1, Top 5, Top 10, and MAP scores for both NetML and AML. This suggests that the SuspWord feature component is more important than the other two (i.e., Text and Spectra).

Furthermore, comparing the results of NetML and AML, we can also observe that the former always gives a better, or at least equal, result than the latter. This suggests that the model parameterization using two sets of model parameters (instead of one in AML), along with the objective function formulation that jointly optimizes bug localization error and fosters clustering of similar bug reports and methods, contribute to a better overall performance of NetML.



\subsubsection{RQ3: Comparisons among Integrator Models}
\label{sec:rq_integrator}

Table~\ref{tab:result_svmrank} compares the performance of the NetML integrator model with that of the AML integrator and SVM$^{rank}$. We can observe that for all projects (i.e., AspectJ, Ant, Lucene, and Rhino) and metrics, the NetML integrator outperforms both the AML integrator and SVM$^{rank}$. With respect to SVM$^{rank}$, NetML achieves 39.76\%, 25.15\%, 20.28\%, and 24.37\% improvements, in terms of Top 1, Top 5, Top 10 and MAP scores across the four software projects, respectively. This can again be attributed to our NetML approach taking advantage of two sets of model parameters and performing a joint optimization of bug localization error and clustering of similar bug reports and methods.

We also conducted the Wilcoxon test to examine whether the improvements over the AML integrator and SVM$^{rank}$ are statistically significant. The resulting $p$-values are summarized in Table \ref{tab:wilcoxon_integrator}. As before, the NetML integrator significantly outperforms the AML integrator in terms of Top 1, Top 5, Top 10, and MAP scores. Moreover, the NetML integrator is significantly better than SVM$^{rank}$ in all evaluation metrics (i.e., Top 1, Top 5, Top 10, and MAP). All in all, these justify the effectiveness of our NetML integrator component.



\begin{table}[t!]
\centering
\caption{Comparisons among different integrator models. The percentage in parentheses indicates the proportion of bug reports whose faulty methods are correctly localized.}
\begin{tabular}{|c|l|c|c|c|}
\hline
\textbf{Metrics}& \textbf{Project}& \textbf{NetML} &\textbf{AML} & \textbf{SVM$^{rank}$}\\
    \hline\hline
\multirow{5}{*}{Top 1} & Ant&13 (24.53\%) &9 (16.98\%)&7 (13.21\%)\\
& AspectJ&11 (26.83\%)&7 (17.07\%)&4 (9.76\%)\\
& Lang&30 (46.15\%)&28 (43.08\%)&27 (41.54\%)\\
& Lucene&12 (32.43\%)&11 (29.73\%)&10 (27.03\%)\\
& Math&32 (30.19\%)&25 (23.58\%)&26(24.53\%)\\
& Rhino&10 (38.46\%)&4 (15.38\%)&4 (15.38\%)\\
& Time&8 (32.77\%)&4 (24.79\%)&5 (23.38\%)\\
\cline{2-5}
&  {\textbf{Overall}}&\textbf{116} (\textbf{32.68\%)}&88 (24.79\%)&83 (23.38\%)\\

\hline
\multirow{5}{*}{Top 5} &Ant&24 (45.28\%)&22 (41.51\%)&24 (45.28\%)\\
& AspectJ&15 (36.59\%)&13 (31.71\%)&11 (26.83\%)\\
&Lang&55 (84.62\%)&48 (73.85\%)&45 (69.23\%)\\
&Lucene&25 (67.57\%)&22 (59.46\%)&23 (62.16\%)\\
&Math&69 (65.09\%)&47 (44.34\%)&46 (43.40\%)\\
&Rhino&18 (69.23\%)&14 (53.85\%)&13 (50.00\%)\\
&Time&13 (48.15\%)&13 (48.15\%)&13 (48.15\%) \\
        \cline{2-5}
		&{\textbf{Overall}}&\textbf{219} \textbf{(61.69\%)}&179 (50.42\%)&175 (49.30\%)\\
\hline
\multirow{5}{*}{Top 10} &     Ant&35 (66.04\%)&31 (58.49\%)&31 (58.49\%)\\
&     AspectJ&16 (39.02\%)&13 (31.71\%)&14 (34.15\%)\\
&    Lang&62 (95.38\%)&53 (81.54\%)&54 (83.08\%)\\
&    Lucene&30 (81.08\%)&29 (78.38\%)&26 (70.27\%)\\
&   Math&75 (70.75\%)&53 (50.00\%)&55 (51.89\%)\\
&   Rhino&19 (73.08\%)&19 (73.08\%)&16 (61.54\%)\\
&   Time&18 (66.67\%)&15 (55.56\%)&16 (59.26\%)\\
\cline{2-5}
&  {\textbf{Overall}}&\textbf{255} \textbf{(71.83\%)}&213 \textbf{(60.00\%)}&212 \textbf{(59.72\%)}\\
\hline
\multirow{5}{*}{MAP}&     Ant&0.270&0.234&0.234\\
&     AspectJ&0.219&0.187&0.131\\
&     Lang&0.638&0.542&0.540\\
&    Lucene&0.290&0.284&0.267\\
&     Math&0.358&0.255&0.269\\
&   Rhino&0.302&0.243&0.227\\
&   Time&0.354&0.294&0.287\\
\cline{2-5}
&  {\textbf{Overall}}&\textbf{0.347}&0.291&0.279\\
\hline
\end{tabular}
\label{tab:result_svmrank}
\end{table}

\begin{table}[!t]
\centering
\caption{The $p$-values of the Wilcoxon test applying the BH procedure on various pairs of integrator model.}
\begin{tabular}{|l|c|c|}
\hline
\textbf{Metrics} & \textbf{NetML vs. SVM$^{rank}$} & \textbf{NetML vs. AML}  \\ 
\hline \hline
\textbf{Top 1} & $ 3 \times 10^{-7}$ (**) & $ 2 \times 10^{-5}$ (**) \\
\textbf{Top 5} & $ 2 \times 10^{-3}$ (**) & $ 1 \times 10^{-3}$ (**) \\
\textbf{Top 10} & $ 4 \times 10^{-4}$ (**) & $ 2 \times 10^{-3}$ (**) \\
\textbf{MAP} & $ 9 \times 10^{-10}$ (**) & $ 2 \times 10^{-8}$ (**) \\
\hline
\multicolumn{3}{l}{(**): smaller than $0.01$}
\end{tabular}
\label{tab:wilcoxon_integrator}
\end{table}

%


\subsubsection{RQ4: Effect of Varying Number of Neighbors}
\label{sec:rq2_neighbor}

%
%


To address this research question, we varied the number of nearest neighbors from $K=5$ to all bug reports in the training set (i.e., $K=\infty$) for both NetML and AML. The results are shown in Table~\ref{tab:varying_neighbors}. We can see that, as we increase $K$, the performance of both multi-modal techniques improves until a certain point (i.e., $K=15$), and decreases beyond that. This suggests that including more neighbors can improve performance to some extent. However, an overly large number of neighbors may lead to an increased level of noise (i.e., the number of irrelevant neighbors), resulting in a degraded performance. Nevertheless, the differences in the Top N and MAP scores are fairly marginal, which justifies the robustness of our NetML approach. Looking at Table~\ref{tab:varying_neighbors}, it is also clear that NetML consistently outperforms AML for all $K$ values (i.e., from $K=5$ to $K=\infty$).




\begin{table*}
	\centering
	\caption{Effect of varying the number of nearest neighbors on NetML and AML. The percentage in parentheses indicates the proportion of bug reports whose faulty methods are correctly localized.}
\begin{tabular}{|l|c|c|c|c|c|c|c|c|}
	\hline 
	\multirow{2}{*}{\textbf{\#Neighbors}} & \multicolumn{2}{c|}{\textbf{Top 1}} & \multicolumn{2}{c|}{\textbf{Top 5}} & \multicolumn{2}{c|}{\textbf{Top 10}} & \multicolumn{2}{c|}{\textbf{MAP}}\\
	\cline{2-9} 
	& \textbf{NetML} & \textbf{AML} & \textbf{NetML} & \textbf{AML} & \textbf{NetML} & \textbf{AML} & \textbf{NetML} & \textbf{AML}\\
	\hline 
	\hline
	$K=5$ & 112 (31.55\%)   & 84 (23.66\%)   & 224 (63.10\%)   & 181 (50.99\%)   & 254 (71.55\%)   & 212 (59.72\%)   & 0.342 & 0.289 \\
	$K=10$ & 116 (32.68\%)  & 88 (24.79\%)   & 219 (61.69\%)   & 179 (50.42\%)  & 255 (71.83\%)   & 213 (60.00\%)  & 0.347 & 0.291 \\
	$K=15$ & 117 (32.96\%)   & 86 (24.23\%)    & 223 (62.82\%)  & 175 (49.30\%)  & 255 (71.83\%)  & 212 (59.72\%)  & 0.347 & 0.291\\
	$K=20$ & 115 (32.39\%)  & 86 (24.23\%)   & 210 (61.97\%)  & 173 (48.73\%)  & 251 (70.70\%)  & 210 (59.15\%)  & 0.345 & 0.29 \\
	$K=25$& 110 (30.99\%)   & 81 (22.81\%)   & 210 (61.97\%)  & 173 (48.73\%)  & 251 (70.70\%)  & 209 (58.87\%)  & 0.331 & 0.285 \\
	$K=\infty$  & 110 (30.99\%)   & 79 (22.26\%)   & 208 (58.59\%)  & 169 (47.61\%)  & 251 (70.70\%)  & 205 (57.75\%)  & 0.329 & 0.283 \\
	\hline
\end{tabular}

\label{tab:varying_neighbors}
\end{table*}

\subsubsection{RQ5: How Effective is NetML in Cross-Project Bug Localization?}
\label{sec:rq5_cross-proj}

\begin{table*}[t!]
	\centering
	\caption{Overall Top N (N $\in {\{1, 5, 10\}}$ and Mean Average Precision (MAP) results in cross-project setting. The percentage in parentheses indicates the proportion of bug reports whose faulty methods are correctly localized.}
	\begin{tabular}{|l|c|c|c|c|}
		\hline
		\textbf{Methods} & \textbf{Top 1}  & \textbf{Top 5}  & \textbf{Top 10} & \textbf{MAP} \\
		\hline
		\hline
		NetML & \textbf{74 (20.85\%)} &\textbf{ 157 (44.23\%)} &\textbf{197 (55.49\%)} & \textbf{0.218} \\
		AML   & 58 (16.34\%) & 131 (36.90\%) & 170 (47.89\%) & 0.174 \\
		Savant & 45 (14.33\%) & 106 (33.76\%) & 135 (42.99\%) & 0.133 \\
		Ochiai & 48 (12.11\%) & 94 (26.48\%) & 124 (34.93\%) & 0.138 \\
		Dstar & 43 (13.52\%) & 88 (24.79\%) & 106 (29.86\%) & 0.114 \\
		\hline
	\end{tabular}%
	\label{tab:cross_proj_sum}%
\end{table*}%

\begin{table*}[t!]
	\centering
	\caption{The $p$-values of the Wilcoxon test applying the BH procedure on various pairs of integrator model in cross-project setting.}
	\begin{tabular}{|l|c|c|c|c|}
		\hline
		\textbf{Method Comparison} & \textbf{Top 1 } & \textbf{Top 5 } & \textbf{Top 10} & \textbf{MAP} \\
		\hline
		\hline
		NetML vs. AML & 0.012 (*) & 0.048 (*) & 0.040 (*)  & 0.029 (*) \\
		NetML vs. Savant & 0.003 (**) & 0.001 (**) & 0.007 (**) & 0.003 (**) \\
		NetML vs. Ochiai & 1 $\times 10^{-5}$ (**) & 0.003 (**) & 8 $\times 10^{-7}$ (**)& 1 $\times 10^{-18}$ (**)\\
		NetML vs. Dstar & $4 \times 10^{-4}$ (**) & 7 $\times 10^{-5}$ (**) & 6 $\times 10^{-7}$ (**) & 7 $\times 10^{-17}$ (**)\\
		\hline
		\multicolumn{5}{l}{(*): smaller than $0.05$, (**): smaller than $0.01$}
	\end{tabular}%
	\label{tab:cross_proj_stats}%
\end{table*}%

\rechecknewagain{Table~\ref{tab:cross_proj_sum} shows the overall performance of NetML and the baseline methods (i.e., AML, Savant, Ochiai, and Dstar) for the cross-project setting, in terms of the Top N and MAP scores respectively. 
Ochiai and Dstar are unsupervised learning methods, which do not depend on training labels. In this case, they give the same result for both cross-project and within-project settings. Hence, we reuse their results in Table~\ref{tab:result_top_N}.
For the remaining techniques (i.e., NetML, AML, and Savant), we use a source project that has the best MAP score for a target project. The results show that NetML outperforms the best baseline (i.e., AML) by 27.59\%, 19.85\%, and 15.88\% in terms of the Top 1, Top 5, and Top 10 methods, respectively. In terms of MAP, NetML outperforms AML, Savant, Ochiai, and Dstar by 25.29\%, 63.91\%, 91.23\%, and 57.97\% respectively. For more details on the Top N results for each pair of source and target projects, please see Tables~\ref{tab:cross_proj_topN} and~\ref{tab:cross_proj_MAP} in the Appendix.}

We also perform Wilcoxon test to compare the overall results of the different techniques in the cross-project setting. Table~\ref{tab:cross_proj_stats} shows the $p$-values for different evaluation metrics (i.e., Top 1, Top 5, Top 10, and MAP) and pairs of techniques. The results indicate that NetML significantly outperforms all the baseline techniques (i.e., AML, Savant, Ochiai, and Dstar) for all the four metrics, thus demonstrating the superior performance of NetML in cross-project setting.

\section{Results Analysis and Discussion}
\label{sec:qualitative}

\rechecknewagain{In this section, we present a detailed analysis of the results obtained in Section~\ref{sec:results}. Firstly, we present some examples to understand the scenarios in which NetML would perform well or poorly in Section~\ref{sec:case_good} and Section~\ref{sec:case_bad}, respectively. Section~\ref{sec:map_across} subsequently presents an analysis on how NetML can improve the MAP performance.}

\begin{figure*}
	\begin{subfigure}{1.0\textwidth}
		\begin{minipage}[!t]{0.5\columnwidth}			
			\begin{tabular}{|p{0.95\columnwidth}|}
				\hline
				\textbf{Program IntrospectionHelper.java}\\ 
				\hline		
				
				
				\textcolor{black}{\textit{public}} void throwNotSupported(\textcolor{black}{\textit{final}}  Project project, \\
				\textcolor{black}{\textit{final}}  Object parent, \textcolor{black}{\textit{final}}  \textcolor{black}{\textit{String}} \textcolor{magenta}{elementName}) \{ \\
				\hspace{2mm} \textcolor{black}{\textit{final}}  \textcolor{black}{\textit{String}} msg = project.\textcolor{cyan}{get}\textcolor{magenta}{ElementName})(parent)  \\ 
				\hspace{2mm} \hspace{2mm} + NOT\_SUPPORTED\_CHILD\_PREFIX \\ 
				\hspace{2mm} \hspace{2mm} + \textcolor{magenta}{elementName}  \\
				\hspace{2mm} \hspace{2mm} + NOT\_SUPPORTED\_CHILD\_POSTFIX; \\
				\hspace{2mm} \textcolor{black}{throw} \textcolor{black}{new} \textcolor{red}{\textit{UnsupportedElementException}}(msg, \textcolor{magenta}{elementName}); \\
				\}
				\\
				\hline
			\end{tabular}
		\end{minipage}\hfill
		\begin{minipage}[!t]{.5\textwidth}
			\begin{tabular}{|p{0.95\columnwidth}|}
				\hline
				\textbf{Program IntrospectionHelper.java}\\ 
				\hline		
				
				\textcolor{black}{\textit{public}} Class$\left\langle?\right\rangle$ getElementType(\textcolor{black}{\textit{final}}  \textcolor{black}{\textit{String}} \textcolor{magenta}{elementName}) throws BuildException \{ \\
				\hspace{2mm} \textcolor{black}{\textit{final}} Class$\left\langle?\right\rangle$ nt = nestedTypes.\textcolor{cyan}{get}(\textcolor{magenta}{elementName}); \\
				\hspace{2mm} \textit{if} (nt == null) \{ \\ 
				\hspace{2mm} \hspace{2mm} \textcolor{black}{throw} \textcolor{black}{new} \textcolor{red}{\textit{UnsupportedElementException}}(``Class'' + \\
				\hspace{2mm} \hspace{2mm} bean.getName() + ``doesn't support the nested $\backslash$'' + \\
				\hspace{2mm} \hspace{2mm} \textcolor{magenta}{elementName} + ``$\backslash$'' element.'', \textcolor{magenta}{elementName});
				\} \\
				\hspace{2mm} return nt; \\
				\}
				\\
				\hline
			\end{tabular}
		\end{minipage}
		\hfill
		\begin{minipage}[!t]{1.0\columnwidth}
			\raggedleft
			\begin{tabular}{|p{0.975\columnwidth}|}
				\hline
				\textbf{Bug 31389}\\ 
				\hline
				\texttt{Summary:}  incorrect error text with invalid ``javac'' task after a ``presetdef''
				\\
				\texttt{Description:} \\
				\textbf{What steps will reproduce the problem?} See below for the build.xml that was used and the faulty error message. \\
				\hspace{2mm} 1. I made a preset definition containing a javac task \\
				\hspace{2mm} 2. I made a normal target (not using the preset definition) containing a javac
				task with an illegal tag name \\
				\hspace{2mm} 3. When running ant, the error message says that the error is in the preset
				definition instead of the javac task. \\ 	\hspace{2mm} (The line number in the message is good.) \\
				\hspace{2mm} \textbf{\dots}
				\\				
				\hline
			\end{tabular}
		\end{minipage}
	\end{subfigure}
	\caption{Example of successful bug localization of two methods in project Ant that need to be resolve the same bug report. The two methods have high cosine similarity score. The colored text indicates some common word tokens occurring in the two methods.}
	\label{fig:method_high_crop}
\end{figure*}

\subsection{Successful Cases}
\label{sec:case_good}

\recheck{We first present two examples of successful bug localization, with the goal of showing how NetML can take advantage of two types of similarity: 1) similarity among bug reports, and 2) similarity among methods.}

\textbf{Bug report similarity}. Our first example involves Bug 30798\footnote{\url{https://bz.apache.org/bugzilla/show_bug.cgi?id=30798}} and Bug 43969\footnote{\url{https://bz.apache.org/bugzilla/show_bug.cgi?id=43969}} from project Ant -- see Fig.~\ref{fig:bug_motivation}. It has been briefly described in Section 1. NetML can outperform AML in identifying the buggy method of Bug 43969 by taking advantage of similarity among bug reports. To confirm that indeed these two bug reports are similar, we can apply the \textit{Vector Space Model} (VSM)~\cite{Manning2008}. We represent each bug report as a TF-IDF vector~\cite{Ramos1999}, and then compute the cosine similarity between the TF-IDF vector of Bug 30798 and that of the remaining bug reports. We find that Bug 43969 is ranked at position \textit{\#}3. Likewise, we compute the cosine similarity between Bug 43969 and the other bug reports. Here, Bug 30798 is ranked at position \textit{\#}5. This shows that Bug 30798 and Bug 43969 are indeed very similar. AML assumes that the bug reports are independent and, owing to the lack of information on the textual description of Bug 43969, it fails to localize the faulty method. In contrast, we found that NetML learns similar model parameters (i.e., $\vec{u}_{b}$) for the two bug reports, and exploits this to compensate for the insufficient information when localizing Bug 43969.

Additionally, we find that none of the other baselines perform as well as NetML. Savant can localize the faulty method of Bug 30798 in its top-10 list, but it fails to do so for Bug 43969. For the other baselines (i.e., Ochiai, Dstar, PROMESIR, DIT$^\text{A}$, DIT$^\text{B}$, LR$^A$, LR$^B$, and MULTRIC), none of them is able to localize the faulty method for both bug reports. Among them, the two best performers (i.e., Ochiai and Dstar) give a high suspiciousness score to the faulty method, but there are more than 100 methods sharing this score.

\textbf{Method similarity}. Fig.~\ref{fig:method_high_crop} presents the description of Bug 31389\footnote{\url{https://bz.apache.org/bugzilla/show_bug.cgi?id=31389}} in project Ant. The bug resides in the \texttt{throwNotSupported} and \texttt{getElementType} methods of \texttt{IntrospectionHelper.java}. NetML is able to localize both methods at positions \textit{\#}1 and \textit{\#}9 respectively, all within the top 10 list. Meanwhile, AML is able to put the \texttt{throwNotSupported} method in the top 10 list, but it ranks the \texttt{getElementType} method at position \textit{\#}17. Ochiai, Dstar, PROMESIR and MULTRIC localize the \texttt{throwNotSupported} method in the top 10 list, but they fail to put the \texttt{getElementType} into the top 10 list. The other baselines give low relevancy scores to the two methods, and exclude them from the top 10 list.

As with the previous example, we try to analyze this further by computing the cosine similarity of the TF-IDF representation of the methods' source code. Specifically, we compute the cosine similarity between \texttt{throwNotSupported} and remaining methods. The result shows that the \texttt{getElementType} method is ranked at position \textit{\#}4. Looking at the content of these two methods, it can again be seen that they share many common word tokens (e.g., ``elementName'', etc.). Accordingly, NetML would enforce the corresponding method parameters to be similar. As such, NetML manages to successfully to localize the \texttt{getElementType} method at position \textit{\#}9. In contrast, AML assumes that the methods are independent, and thus fails to leverage the strength of similar methods to localize the \texttt{getElementType} method.

\begin{figure*}
	\begin{subfigure}{1.0\textwidth}
		\begin{minipage}[!t]{0.5\columnwidth}			
			\begin{tabular}{|p{0.95\columnwidth}|}
				\hline
				\textbf{Bug 338}\\ 
				\hline
				\texttt{Summary:} 
				Wrong parameter for first step size guess for Embedded Runge Kutta methods
				\\    
				\\
				\texttt{Description:}
				\\
				\textbf{What steps will reproduce the problem?} In a space application using DOP853 i detected what seems to be a bad parameter in the call to the method initializeStep of class AdaptiveStepsizeIntegrator.  \dots 
				\\
				\hline
			\end{tabular}
		\end{minipage}\hfill
		\begin{minipage}[!t]{.5\textwidth}
			\begin{tabular}{|p{0.95\columnwidth}|}
				\hline
				\textbf{Bug 358}\\ 
				\hline
				\texttt{Summary:} 
				ODE integrator goes past specified end of integration range
				
				\\
				\texttt{Description:}
				\\
				\textbf{What steps will reproduce the problem?} End of integration range in ODE solving is handled as an event. In some cases, numerical accuracy in events detection leads to error in events location. \dots
				\\				
				\hline
			\end{tabular}
		\end{minipage}\hfill
	\end{subfigure}\hfill
	\caption{Example of unsuccessful bug localization of two bug reports which have the same faulty method in project Math. The two bug reports have low cosine similarity score}
	\label{fig:bug_low}
\end{figure*}

\begin{figure*}[!t]
	\centering
	\begin{subfigure}{1.0\textwidth}
		\begin{minipage}[!t]{0.5\columnwidth}			
			\begin{tabular}{|p{0.95\columnwidth}|}
				\hline
				\textbf{Program ChangeLogParser.java}\\ 
				\hline				
				\textit{private} Date parseDate(\textit{final} \textit{String} date) \{ \\
				\hspace{2mm} try \{ \\
				\hspace{2mm} \hspace{2mm} return c\_inputDate.parse(date); \\
				\hspace{2mm} \} catch (ParseException e) \{ \\
				\hspace{2mm} \hspace{2mm} //\textit{final} \textit{String} message = REZ.getString(\\
				\hspace{2mm} \hspace{2mm} //``changelog.bat-date.error'', date); \\
				\hspace{2mm} \hspace{2mm} //getContext().error( message ); \\
				\hspace{2mm} \hspace{2mm} return null; \\
				\hspace{2mm} \hspace{2mm} \} \\
				\}
				\\
				\\
				\hline
			\end{tabular}
		\end{minipage}\hfill
		\begin{minipage}[!t]{.5\textwidth}
			\begin{tabular}{|p{0.95\columnwidth}|}
				\hline
				\textbf{Program ChangeLogParser.java}\\ 
				\hline		
				\textit{private} void processGetPreviousRevision(\textit{final} \textit{String} line)\\ 
				\{ \\
				\hspace{2mm} \textit{if} (!line.startsWith(``revision''))\{ \\
				\hspace{2mm} \hspace{2mm} throw new IllegalStateException(``Unexpected line \\ 
				\hspace{2mm} \hspace{2mm} from CVS:'' + line); \\ 
				\hspace{2mm} \} \\
				\hspace{2mm} m\_previousRevision = line.substring(9); \\
				\hspace{2mm} saveEntry(); \\
				\hspace{2mm} m\_revision = m\_previousRevision; \\
				\hspace{2mm} m\_status = GET\_DATE; \\
				\}
				\\
				\hline
			\end{tabular}
		\end{minipage} \hfill
		\begin{minipage}[!t]{1.0\columnwidth}
			\raggedleft
			\begin{tabular}{|p{0.975\columnwidth}|}
				\hline
				\textbf{Bug 30962}\\ 
				\hline
				\texttt{Summary:} 
				cvschangelog crashes with NullPointerException
				\\
				\texttt{Description:}
				
				\textbf{What steps will reproduce the problem?} I try to make cvschangelog running and face a strange problem that nobody else seems to have: cvschangelog crashes with a NullPointerException. My task looks like: \\
				\hspace{2mm} $\langle$target name=``cvs.changelog''$\rangle$ \\
				\hspace{2mm} $\langle$cvschangelog dir=``somedir'' destfile=``changelog.xml''$\rangle$  \\				
				\hspace{2mm} \textbf{\dots}		
				\\				
				\hline
			\end{tabular}
		\end{minipage}\hfill
		
	\end{subfigure}\hfill
	\caption{Example of unsuccessful bug localization of two methods in project Ant that need to be modified to resolve the same bug report. The two methods have low cosine similarity score. 
	}
	\label{fig:method_low}
\end{figure*}

\rechecknewagain{To see how typical the successful cases are in our dataset, we randomly select
75 out of 183 successful cases, in which NetML manages to localize a faulty method within the top 10 list whereas the other baseline methods (i.e., AML, Savant, Ochiai, Dstar, PROMESIR, DIT$^\text{A}$, DIT$^\text{B}$, LR$^A$, LR$^B$, and MULTRIC) 
fail to do so. Among these cases, in total, we find that 63 successful cases, which constitute the majority (84\%) of our samples, are similar to the first (17 cases) and second (46 cases) examples we presented earlier.
}

\subsection{Unsuccessful Cases}
\label{sec:case_bad}


Next, we present two examples whereby NetML fails to localize a bug. These examples serve to provide an understanding of cases in which NetML may not perform well.

\textbf{Bug report similarity}. We first consider Bug 338\footnote{\url{https://issues.apache.org/jira/browse/MATH-338}} and Bug 358\footnote{\url{https://issues.apache.org/jira/browse/MATH-358}} from project Math shown in Fig.~\ref{fig:bug_low}. The faulty method for these two bug reports is the \texttt{integrate} method in \texttt{EmbeddedRungeKuttaIntegrator.java}. Interestingly, Ochiai and Dstar manage to localize this faulty method for these two bug reports within the top 10 list. On the other hand, NetML, AML, and Savant fail to localize the faulty \texttt{integrate} method for Bug 358. Specifically, NetML, AML, and Savant rank the faulty method at positions \textit{\#}14, \textit{\#}19, and \textit{\#}23 respectively. MULTRIC assigns a high suspiciousness score to the \texttt{integrate} method for both Bug 338 and Bug 358. However, there are around 30 methods sharing this score. Also note that the remaining baselines (i.e., PROMESIR, DIT$^\text{A}$, DIT$^\text{B}$, LR$^A$, and LR$^B$) fail to localize the faulty method for both bug reports.

Similar to Section~\ref{sec:case_good}, we calculate the cosine similarity between Bug 338 and the remaining bug reports. We found that Bug 358 is ranked at position \textit{\#}53, suggesting that the two bug reports are dissimilar. As such, there is less incentive for NetML to leverage the strength of common words shared by the two bug reports, which potentially explains why it fails to localize the faulty method for Bug 358. This also suggests that, when the data contain bug reports that are largely dissimilar (i.e., share very few common word tokens), our NetML approach may not work as well as some spectrum-based fault localization techniques such as Ochiai and Dstar.

\textbf{Method similarity}. Fig.~\ref{fig:method_low} shows the descriptions of Bug 30962\footnote{\url{https://bz.apache.org/bugzilla/show_bug.cgi?id=30962}} in project Ant. The bug is associated with two faulty methods, i.e., \texttt{parseDate} and \texttt{processGetPreviousRevision} in  \texttt{ChangeLogParser.java}. We find that Ochiai and Dstar successfully localize these two methods in the top 10 list. NetML and AML are able to localize the \texttt{parseDate} method within the top 10 list. However, they fail to localize the faulty \texttt{processGetPreviousRevision} method for Bug 30962. In particular, NetML and AML place the \texttt{processGetPreviousRevision} method at positions \textit{\#}17 and \textit{\#}15 respectively. The remaining techniques (i.e., Savant, PROMESIR, DIT$^\text{A}$, DIT$^\text{B}$, LR$^A$, LR$^B$) fail to localize these two methods in the top 10 list.

To better understand this, we again compute the cosine similarity between the \texttt{parseDate} method and the remaining methods in project Ant. In this case, the \texttt{processGetPreviousRevision} method is ranked at position \textit{\#}478. This suggests that these two methods have low proximity, which gives less incentive for NetML to utilize their common words in the localization of the \texttt{processGetPreviousRevision} method. It also suggests that, when the data contain methods that are mostly dissimilar, spectrum-based fault localization techniques (e.g., Ochiai and Dstar) may perform better than NetML.


\rechecknewagain{To again evaluate how typical the unsuccessful cases are in seven projects, we randomly select 75 (out of 80) unsuccessful cases whereby NetML fails to localize a faulty method within the top 10 list, but one of the baseline method (i.e., AML, Savant, Ochiai, Dstar, PROMESIR, DIT$^\text{A}$, DIT$^\text{B}$, LR$^A$, LR$^B$, and MULTRIC) succeed. Among them, in total, we discover that 70 unsuccessful cases, which constitute 93\% of our samples, are similar to the first (21 cases) and second (49 cases) unsuccessful examples presented earlier.}

\subsection{Improved vs. Deteriorated Bug Reports}
\label{sec:map_across}

\begin{table*}[t!]
	\centering
	\caption{Comparison of number of samples, expected average precision difference, and expected rank difference between NetML and AML.}
	\begin{tabular}{|l|c|c|c|c|c|c|c|}
		\hline
		\multicolumn{1}{|c|}{\multirow{2}[4]{*}{\textbf{Project}}} & \multicolumn{3}{c|}{\textbf{Improved}} & \multicolumn{3}{c|}{\textbf{Deteriorated}} & \multicolumn{1}{c|}{\multirow{1}[2]{*}{\textbf{Unchanged}}} \\
		\cline{2-8}          & \textbf{No. of samples} & \textbf{$E[\Delta{AP}]$} & \textbf{$E[\Delta{Rank}]$} & \textbf{No. of samples} & \textbf{$E[\Delta{AP}]$} & \textbf{$E[\Delta{Rank}]$} &  \textbf{No. of samples}\\
		\hline
		\hline
		Ant   & 35 (66.04\%) & 12.57\% & 186.86 & 3 (5.66\%) & -8.47\% & -54.27 & 15 (28.3\%) \\
		\hline
		Aspectj & 32 (78.05\%) & 23.02\% & 59.31 & 3 (7.32\%) & -39.67\% & -39.67 & 6 (14.63\%) \\
		\hline
		Lang  & 37 (56.92\%) & 29.94\% & 26.93 & 5 (7.69\%) & -36.11\% & -35.52 & 23 (35.38\%) \\
		\hline
		Lucene & 14 (37.84\%) & 11.94\% & 372.64 & 10 (27.03\%) & -14.97\% & -297.2 & 13 (35.14\%) \\
		\hline
		Math  & 70 (66.04\%) & 37.37\% & 16.45 & 10 (9.43\%) & -6.33\% & -13.07 & 26 (24.53\%) \\
		\hline
		Rhino & 22 (84.62\%) & 36.39\% & 54.54 & 2 (7.69\%) & -15.51\% & -15.32 & 2 (7.69\%) \\
		\hline
		Time  & 16 (59.26\%) & 24.15\% & 60.5  & 6 (22.22\%) & -7.54\% & -10.12 & 5 (18.52\%) \\
		\hline
		\textbf{Overall} & \textbf{229 (63.66\%)} &\textbf{27.75\%} &\textbf{212.91} & \textbf{39 (10.98\%)} &\textbf{-14.36\%} & \textbf{-69.09} & \textbf{90 (25.35\%)} \\
		\hline
	\end{tabular}%
	\label{tab:map_improve}%
\end{table*}%

\rechecknewagain{To understand how the MAP results improve due to NetML, following Chaparro et al.~\cite{chaparro2017using}, we perform a finer-grained analysis in terms of the number of bug reports improved/deteriorated and the expected magnitude of improvement/deterioration. We compare our approach against the best baseline method (i.e., AML). A bug report is \textit{improved} if the rank of the top faulty method produced by NetML is better than the rank of the top faulty method produced by AML. On the other hand, a bug report is \textit{deteriorated} if the rank of the top faulty method produced by NetML is worse than that produced by AML. Otherwise, a bug report is \textit{unchanged}. Ideally, we wish to have a higher number of \textit{improved} bug reports than that of \textit{deteriorated} bug reports. To measure the relative magnitude of improvement or deterioration for each bug report, we adopt the approach described in~\cite{chaparro2017using}. In particular, for \textit{improved} and \textit{deteriorated} bug reports, we compute the expected average precision (AP) difference $E[\Delta AP]$ and expected rank difference $E[\Delta Rank]$ as follows:}

\begin{equation}
	E[\Delta AP] = \dfrac{1}{|\mathcal{B}|} \sum_{b=1}^{\mathcal{B}} (AP_b^{NetML} - AP_b^{AML})
\label{eq:MAP_diff}
\end{equation}

\begin{equation}
E[\Delta Rank] = \dfrac{1}{|\mathcal{B}|} \sum_{b=1}^{\mathcal{B}} (Rank_b^{AML} - Rank_b^{NetML})
\label{eq:Rank_diff}
\end{equation}
\rechecknewagain{where $|\mathcal{B}|$ is the number of bug reports, $AP_b^{NetML}$ and $AP_b^{AML}$ are the average precision produced by NetML and AML for bug report $b$, and $Rank_b^{NetML}$ and $Rank_b^{AML}$ are the rank produced by NetML and AML, respectively. Intuitively, if NetML is better than AML, we expect the $E[\Delta AP]$ and $E[\Delta Rank]$ for \textit{improved} bug reports to be larger than those of \textit{deteriorated} bug reports.}

\rechecknewagain{Table~\ref{tab:map_improve} shows the number of \textit{improved}, \textit{deteriorated}, and \textit{unchanged} bug reports in our seven projects. Additionally, Table~\ref{tab:map_improve} presents the $E[\Delta AP]$ and $E[\Delta Rank]$ for \textit{improved} and \textit{deteriorated} cases of different projects. 
The results show that the number of \textit{improved} bug reports is indeed higher than the number of \textit{deteriorated} bug reports for all different projects.
It is also evident that the overall $E[\Delta AP]$ and $E[\Delta Rank]$ of \textit{improved} bug reports are higher than those of \textit{deteriorated} bug reports. This implies that MAP improvement comes from improvements across the boards and not due to a few outlier bug reports or projects.}

\section{Threats to Validity}
\label{sec:threats}

This section presents a number of threats that may potentially impact the validity of our study.

\subsection{Number of Failed Test Cases and Its Impact}

In our experiments with 355 bugs, most of the bugs were found to come with few failed test cases (average = 2.155). We have investigated if the number of failed test cases impacts the effectiveness of our approach. To this end, we computed the differences between the average number of failed test cases for bugs that are successfully localized at Top-N positions (N = 1,5,10) and bugs that are not successfully localized. We found that the differences are small (-0.362 to 0.055 test cases). These indicate that the number of test cases does not impact the effectiveness of our approach significantly and typically 1 to 3 failed test cases are sufficient for our approach to be effective.

\subsection{Threats to Internal Validity}

Threats to internal validity relate to implementation and dataset errors. We have checked our implementations and datasets. However, there could still be errors that we do not notice. Threats to external validity relate to the generalizability of our findings. In this work, we have analyzed 355 real bugs from seven medium-large software systems. In the future, we plan to reduce the threats to external validity by investigating more real bugs from additional software systems, written in various programming languages.

\section{Related Work}
\label{sec.related}

In this section, we highlight a number of research studies that are closely related to our work.

\begin{table*}[t!]
	\centering
	\caption{\recheckagain{Top N (N $\in \{1, 5, 10\}$) results of NetML vs. AML, Savant, Ochiai, Dstar, and PROMESIR. The percentage in parentheses indicates the proportion of bug reports whose faulty methods are correctly localized}.}
	\begin{tabular}{|c|l|c|c|c|c|c|c|}
		\hline
		\textbf{Top N} & \multicolumn{1}{c|}{\textbf{Project}} & \textbf{NetML} & \textbf{AML} & \textbf{Savant} & \textbf{Ochiai} & \textbf{Dstar} & \textbf{PROMESIR} \\
		\hline
		\hline
		\multirow{8}[2]{*}{1} & Ant   & 13 (24.53\%) & 9 (16.98\%) & 8 (15.09\%) & 6 (11.32\%) & 6 (11.32\%) & 7 (13.21\%) \\
		& Aspectj & 11 (26.83\%) & 7 (17.07\%) & --    & 3 (7.32\%) & 1 (2.44\%) & 4 (9.76\%) \\
		& Lang  & 30 (46.15\%) & 28 (43.08\%) & 29 (44.62\%) & 13 (20.00\%) & 12 (18.46\%) & 19 (29.23\%) \\
		& Lucene & 12 (32.43\%) & 11 (29.73\%) & 3 (8.11\%) & 6 (16.22\%) & 5 (13.51\%) & 8 (21.62\%) \\
		& Math  & 32 (30.19\%) & 25 (23.58\%) & 22 (20.75\%) & 14 (13.21\%) & 13 (12.26\%) & 20 (18.87\%) \\
		& Rhino & 10 (38.46\%) & 4 (15.38\%) & 0 (0.00\%) & 4 (15.38\%) & 4 (15.38\%) & 2 (7.69\%) \\
		& Time  & 8 (29.63\%) & 4 (14.81\%) & 5 (18.52\%) & 2 (7.41\%) & 2 (7.41\%) & 1 (3.70\%) \\
		\cline{2-8}
		& \textbf{Overall} & \textbf{116 (32.68\%)} & 88 (24.79\%) & 67 (21.34\%) & 48 (13.52\%) & 43 (12.11\%) & 61 (17.18\%) \\
		\hline
		\multirow{8}[2]{*}{5} & Ant   & 24 (45.28\%) & 22 (41.51\%) & 11 (20.75\%) & 12 (22.64\%) & 10 (18.87\%) & 17 (32.08\%) \\
		& Aspectj & 15 (36.59\%) & 13 (31.71\%) & --    & 5 (12.20\%) & 1 (2.44\%) & 6 (14.63\%) \\
		& Lang  & 55 (84.62\%) & 48 (73.85\%) & 41 (63.08\%) & 20 (30.77\%) & 22 (33.85\%) & 38 (58.46\%) \\
		& Lucene & 25 (67.57\%) & 22 (59.46\%) & 7 (18.92\%) & 14 (37.84\%) & 14 (37.84\%) & 18 (48.65\%) \\
		& Math  & 69 (65.09\%) & 47 (44.34\%) & 47 (44.34\%) & 30 (28.30\%) & 30 (28.30\%) & 44 (41.51\%) \\
		& Rhino & 18 (69.23\%) & 14 (53.85\%) & 4 (15.38\%) & 8 (30.77\%) & 6 (23.08\%) & 13 (50.00\%) \\
		& Time  & 13 (48.15\%) & 13 (48.15\%) & 12 (44.44\%) & 5 (18.52\%) & 5 (18.52\%) & 3 (11.11\%) \\
		\cline{2-8}
		& \textbf{Overall} & \textbf{219 (61.69\%)} & 179 (50.42\%) & 122 (38.85\%) & 94 (26.48\%) & 88 (24.79\%) & 139 (39.15\%) \\
		\hline
		\multirow{8}[2]{*}{10} & Ant   & 35 (66.04\%) & 31 (58.49\%) & 15 (28.3\%) & 23 (43.40\%) & 16 (30.19\%) & 28 (52.83\%) \\
		& Aspectj & 16 (39.02\%) & 13 (31.71\%) & --    & 6 (14.63\%) & 1 (2.44\%) & 9 (21.95\%) \\
		& Lang  & 62 (95.38\%) & 53 (81.54\%) & 45 (69.23\%) & 26 (40.00\%) & 25 (38.46\%) & 46 (70.77\%) \\
		& Lucene & 30 (81.08\%) & 29 (78.38\%) & 12 (32.43\%) & 15 (40.54\%) & 15 (40.54\%) & 21 (56.76\%) \\
		& Math  & 75 (70.75\%) & 53 (50\%) & 57 (53.77\%) & 39 (36.79\%) & 38 (35.85\%) & 52 (49.06\%) \\
		& Rhino & 19 (73.08\%) & 19 (73.08\%) & 9 (34.62\%) & 9 (34.62\%) & 6 (23.08\%) & 14 (53.85\%) \\
		& Time  & 18 (66.67\%) & 15 (55.56\%) & 14 (51.85\%) & 6 (22.22\%) & 5 (18.52\%) & 4 (14.81\%) \\
		\cline{2-8}
		& \textbf{Overall} & \textbf{255 (71.83\%)} & 213 (60.00\%) & 152 (48.41\%) & 124 (34.93\%) & 106 (29.86\%) & 174 (49.01\%) \\
		\hline
	\end{tabular}%
	\label{tab:full_result_top_N}%
\end{table*}%

\begin{table*}[t!]
	\centering
	\caption{\recheckagain{Top N (N $\in \{1, 5, 10\}$) results of NetML vs. DIT$^\text{A}$, DIT$^\text{B}$, LR$^A$, LR$^B$, and MULTRIC. The percentage in parentheses indicates the proportion of bug reports whose faulty methods are correctly localized}.}
	\begin{tabular}{|c|l|c|c|c|c|c|c|}
		\hline
		\textbf{Top N} & \multicolumn{1}{c|}{\textbf{Project}} & \textbf{NetML} & \textbf{DIT$^\text{A}$} & \textbf{DIT$^\text{B}$} & \textbf{LR$^A$} & \textbf{LR$^B$} & \textbf{MULTRIC} \\
		\hline
		\hline
		\multirow{8}[2]{*}{1} & Ant   & 13 (24.53\%) & 3 (5.66\%) & 3 (5.66\%) & 1 (1.89\%) & 11 (20.75\%) & 2 (3.77\%) \\
		& Aspectj & 11 (26.83\%) & 4 (9.76\%) & 3 (7.32\%) & 0 (0.00\%) & 0 (0.00\%) & 0 (0.00\%) \\
		& Lang  & 30 (46.15\%) & 12 (18.46\%) & 11 (16.92\%) & 4 (6.15\%) & 21 (32.31\%) & 23 (35.38\%) \\
		& Lucene & 12 (32.43\%) & 7 (18.92\%) & 7 (18.92\%) & 1 (2.70\%) & 7 (18.92\%) & 4 (10.81\%) \\
		& Math  & 32 (30.19\%) & 13 (12.26\%) & 12 (11.32\%) & 6 (5.66\%) & 23 (21.70\%) & 30 (28.3\%) \\
		& Rhino & 10 (38.46\%) & 2 (7.69\%) & 1 (3.85\%) & 0 (0.00\%) & 2 (7.69\%) & 2 (7.69\%) \\
		& Time  & 8 (29.63\%) & 0 (0.00\%) & 0 (0.00\%) & 0 (0.00\%) & 2 (7.41\%) & 7 (25.93\%) \\
		\cline{2-8}
		& \textbf{Overall} & \textbf{116 (32.68\%)} & 41 (11.55\%) & 37 (10.42\%) & 12 (3.38\%) & 66 (18.59\%) & 68 (19.15\%) \\
		\hline
		\multirow{8}[2]{*}{5} & Ant   & 24 (45.28\%) & 10 (18.87\%) & 10 (18.87\%) & 11 (20.75\%) & 20 (37.74\%) & 7 (13.21\%) \\
		& Aspectj & 15 (36.59\%) & 4 (9.76\%) & 3 (7.32\%) & 0 (0.00\%) & 0 (0.00\%) & 1 (2.44\%) \\
		& Lang  & 55 (84.62\%) & 17 (26.15\%) & 17 (26.15\%) & 13 (20.00\%) & 39 (60.00\%) & 39 (60.00\%) \\
		& Lucene & 25 (67.57\%) & 13 (35.14\%) & 13 (35.14\%) & 6 (16.22\%) & 16 (43.24\%) & 13 (35.14\%) \\
		& Math  & 69 (65.09\%) & 30 (28.30\%) & 29 (27.36\%) & 34 (32.08\%) & 48 (45.28\%) & 55 (51.89\%) \\
		& Rhino & 18 (69.23\%) & 13 (50.00\%) & 5 (19.23\%) & 2 (7.69\%) & 8 (30.77\%) & 8 (30.77\%) \\
		& Time  & 13 (48.15\%) & 1 (3.70\%) & 1 (3.70\%) & 1 (3.70\%) & 6 (22.22\%) & 10 (37.04\%) \\
		\cline{2-8}
		& \textbf{Overall} & \textbf{219 (61.69\%)} & 88 (24.79\%) & 78 (21.97\%) & 67 (18.87\%) & 137 (38.59\%) & 133 (37.46\%) \\
		\hline
		\multirow{8}[2]{*}{10} & Ant   & 35 (66.04\%) & 20 (37.74\%) & 20 (37.74\%) & 19 (35.85\%) & 32 (60.38\%) & 15 (28.30\%) \\
		& Aspectj & 16 (39.02\%) & 4 (9.76\%) & 3 (7.32\%) & 0 (0.00\%) & 0 (0.00\%) & 2 (4.88\%) \\
		& Lang  & 62 (95.38\%) & 23 (35.38\%) & 22 (33.85\%) & 30 (46.15\%) & 46 (70.77\%) & 41 (63.08\%) \\
		& Lucene & 30 (81.08\%) & 20 (54.05\%) & 20 (54.05\%) & 10 (27.03\%) & 24 (64.86\%) & 16 (43.24\%) \\
		& Math  & 75 (70.75\%) & 33 (31.13\%) & 34 (32.08\%) & 51 (48.11\%) & 59 (55.66\%) & 65 (61.32\%) \\
		& Rhino & 19 (73.08\%) & 14 (53.85\%) & 7 (26.92\%) & 3 (11.54\%) & 12 (46.15\%) & 11 (42.31\%) \\
		& Time  & 18 (66.67\%) & 3 (11.11\%) & 3 (11.11\%) & 3 (11.11\%) & 8 (29.63\%) & 12 (44.44\%) \\
		\cline{2-8}
		& \textbf{Overall} & \textbf{255 (71.83\%)} & 117 (32.96\%) & 109 (30.7\%) & 116 (32.68\%) & 181 (50.99\%) & 162 (45.63\%) \\
		\hline
	\end{tabular}%
	\label{tab:full_result_top_N_}%
\end{table*}%

\begin{table*}[t!]
	\centering
	\caption{Top N (N $\in {\{1, 5, 10\}}$ results on cross-project setting, for different pairs of source and target projects. The percentage in parentheses indicates the proportion of bug reports whose faulty methods are correctly localized.}
	\begin{adjustbox} {width=1\textwidth} 
		\begin{tabular}{|l|l|c|c|c|c|c|c|c|c|c|}
			\hline
			\multirow{2}[4]{*}{\textbf{Source}} & \multirow{2}[4]{*}{\textbf{Target}} & \multicolumn{3}{c|}{\textbf{Top 1}} & \multicolumn{3}{c}{\textbf{Top 5}} & \multicolumn{3}{c|}{\textbf{Top 10}} \\
			\cline{3-11}          &       & \textbf{NetML} & \textbf{AML} & \textbf{Savant} & \textbf{NetML} & \textbf{AML} & \textbf{Savant} & \textbf{NetML} & \textbf{AML} & \textbf{Savant} \\
			\hline
			\hline
			Aspectj & Ant   & 8 (15.09\%) & 8 (15.09\%) & --     & 20 (37.74\%) & 20 (37.74\%) & --    & 28 (58.49\%) & 31 (52.83\%) & - \\
			Lang  & Ant   & 8 (15.09\%) & 7 (13.21\%) & 6 (11.32\%) & 19 (39.62\%) & 21 (35.85\%) & 9 (16.98\%) & 28 (56.6\%) & 30 (52.83\%) & 12 (22.64\%) \\
			Lucene & Ant   & 8 (15.09\%) & 7 (13.21\%) & 5 (9.43\%) & 17 (43.4\%) & 23 (32.08\%) & 11 (20.75\%) & 26 (58.49\%) & 31 (49.06\%) & 15 (28.30\%) \\
			Math  & Ant   & 9 (16.98\%) & 8 (15.09\%) & 4 (7.55\%) & 24 (41.51\%) & 22 (45.28\%) & 10 (18.87\%) & 31 (58.49\%) & 31 (58.49\%) & 13 (24.53\%) \\
			Rhino & Ant   & 8 (15.09\%) & 7 (13.21\%) & 7 (13.21\%) & 21 (39.62\%) & 21 (39.62\%) & 9 (16.98\%) & 27 (56.60\%) & 30 (50.94\%) & 14 (26.42\%) \\
			Time  & Ant   & 7 (13.21\%) & 7 (13.21\%) & 2 (3.77\%) & 21 (28.30\%) & 15 (39.62\%) & 3 (5.66\%) & 30 (50.94\%) & 27 (56.60\%) & 3 (5.66\%) \\
			\hline
			Ant   & Aspectj & 2 (4.88\%) & 3 (7.32\%) & \multirow{6}[2]{*}{--} & 7 (17.07\%) & 9 (21.95\%) & \multirow{6}[2]{*}{-} & 9 (21.95\%) & 9 (21.95\%) & \multirow{6}[2]{*}{-} \\
			Lang  & Aspectj & 3 (7.32\%) & 3 (7.32\%) &       & 9 (21.95\%) & 8 (19.51\%) &       & 10 (24.39\%) & 9 (21.95\%) &  \\
			Lucene & Aspectj & 4 (9.76\%) & 3 (7.32\%) &       & 8 (19.51\%) & 8 (19.51\%) &       & 10 (24.39\%) & 9 (21.95\%) &  \\
			Math  & Aspectj & 3 (7.32\%) & 2 (4.88\%) &       & 11 (26.83\%) & 3 (7.32\%) &       & 11 (26.83\%) & 4 (9.76\%) &  \\
			Rhino & Aspectj & 4 (9.76\%) & 4 (9.76\%) &       & 10 (24.39\%) & 6 (14.63\%) &       & 11 (26.83\%) & 8 (19.51\%) &  \\
			Time  & Aspectj & 7 (17.07\%) & 2 (4.88\%) &       & 11 (26.83\%) & 3 (7.32\%) &       & 13 (31.71\%) & 5 (12.20\%) &  \\
			\hline
			Ant   & Lang  & 17 (26.15\%) & 17 (26.15\%) & 7 (10.77\%) & 35 (53.85\%) & 31 (47.69\%) & 29 (44.62\%) & 45 (69.23\%) & 39 (60.00\%) & 37 (56.92\%) \\
			Aspectj & Lang  & 17 (26.15\%) & 16 (24.62\%) & --     & 38 (58.46\%) & 31 (47.69\%) & --     & 46 (70.77\%) & 37 (56.92\%) & - \\
			Lucene & Lang  & 15 (23.08\%) & 17 (26.15\%) & 7 (10.77\%) & 34 (52.31\%) & 33 (50.77\%) & 26 (40.00\%) & 42 (64.62\%) & 40 (61.54\%) & 38 (58.46\%) \\
			Math  & Lang  & 19 (29.23\%) & 17 (26.15\%) & 17 (26.15\%) & 36 (55.38\%) & 31 (47.69\%) & 38 (58.46\%) & 42 (64.62\%) & 39 (60.00\%) & 44 (67.69\%) \\
			Rhino & Lang  & 19 (29.23\%) & 17 (26.15\%) & 10 (15.38\%) & 38 (58.46\%) & 31 (47.69\%) & 37 (56.92\%) & 45 (69.23\%) & 38 (58.46\%) & 43 (66.15\%) \\
			Time  & Lang  & 20 (30.77\%) & 12 (18.46\%) & 15 (23.08\%) & 39 (60.00\%) & 29 (44.62\%) & 38 (58.46\%) & 46 (70.77\%) & 35 (53.85\%) & 46 (70.77\%) \\
			\hline
			Aspectj & Lucene & 7 (18.92\%) & 7 (18.92\%) & --     & 19 (51.35\%) & 17 (45.95\%) & --     & 23 (62.16\%) & 22 (59.46\%) & - \\
			Ant   & Lucene & 9 (24.32\%) & 7 (18.92\%) & 2 (5.41\%) & 22 (59.46\%) & 16 (43.24\%) & 7 (18.92\%) & 27 (72.97\%) & 21 (56.76\%) & 7 (18.92\%) \\
			Lang  & Lucene & 9 (24.32\%) & 7 (18.92\%) & 2 (5.41\%) & 20 (54.05\%) & 16 (43.24\%) & 6 (16.22\%) & 25 (67.57\%) & 21 (56.76\%) & 9 (24.32\%) \\
			Math  & Lucene & 8 (21.62\%) & 7 (18.92\%) & 4 (10.81\%) & 21 (56.76\%) & 16 (43.24\%) & 7 (18.92\%) & 25 (67.57\%) & 21 (56.76\%) & 9 (24.32\%) \\
			Rhino & Lucene & 10 (27.03\%) & 7 (18.92\%) & 9 (24.32\%) & 23 (62.16\%) & 15 (40.54\%) & 16 (43.24\%) & 26 (70.27\%) & 21 (56.76\%) & 17 (45.95\%) \\
			Time  & Lucene & 10 (27.03\%) & 6 (16.22\%) & 2 (5.41\%) & 26 (70.27\%) & 12 (32.43\%) & 3 (8.11\%) & 28 (75.68\%) & 16 (43.24\%) & 3 (8.11\%) \\
			\hline
			Ant   & Math  & 20 (18.87\%) & 14 (13.21\%) & 6 (5.66\%) & 39 (36.79\%) & 39 (36.79\%) & 8 (7.55\%) & 54 (50.94\%) & 49 (46.23\%) & 12 (11.32\%) \\
			Aspectj & Math  & 19 (17.92\%) & 14 (13.21\%) & --     & 42 (39.62\%) & 35 (33.02\%) & --     & 50 (47.17\%) & 49 (46.23\%) & - \\
			Lang  & Math  & 17 (16.04\%) & 14 (13.21\%) & 8 (7.55\%) & 41 (38.68\%) & 37 (34.91\%) & 15 (14.15\%) & 52 (49.06\%) & 48 (45.28\%) & 24 (22.64\%) \\
			Lucene & Math  & 16 (15.09\%) & 14 (13.21\%) & 6 (5.66\%) & 36 (33.96\%) & 38 (35.85\%) & 15 (14.15\%) & 47 (44.34\%) & 49 (46.23\%) & 17 (16.04\%) \\
			Rhino & Math  & 16 (15.09\%) & 14 (13.21\%) & 9 (8.49\%) & 42 (39.62\%) & 38 (35.85\%) & 20 (18.87\%) & 53 (50.00\%) & 49 (46.23\%) & 30 (28.30\%) \\
			Time  & Math  & 15 (14.15\%) & 11 (10.38\%) & 11 (10.38\%) & 42 (39.62\%) & 35 (33.02\%) & 36 (33.96\%) & 53 (50.00\%) & 49 (46.23\%) & 50 (47.17\%) \\
			\hline
			Ant   & Rhino & 4 (15.38\%) & 3 (11.54\%) & 0 (0.00\%) & 11 (42.31\%) & 11 (42.31\%) & 1 (3.85\%) & 15 (57.69\%) & 14 (53.85\%) & 1 (3.85\%) \\
			Aspectj & Rhino & 5 (19.23\%) & 5 (19.23\%) & --     & 12 (46.15\%) & 10 (38.46\%) & --     & 17 (65.38\%) & 14 (53.85\%) & -- \\
			Lang  & Rhino & 4 (15.38\%) & 2 (7.69\%) & 2 (7.69\%) & 9 (34.62\%) & 8 (30.77\%) & 2 (7.69\%) & 11 (42.31\%) & 13 (50.00\%) & 4 (15.38\%) \\
			Lucene & Rhino & 4 (15.38\%) & 3 (11.54\%) & 0 (0.00\%) & 7 (26.92\%) & 11 (42.31\%) & 1 (3.85\%) & 8 (30.77\%) & 15 (57.69\%) & 2 (7.69\%) \\
			Math  & Rhino & 5 (19.23\%) & 3 (11.54\%) & 2 (7.69\%) & 12 (46.15\%) & 10 (38.46\%) & 2 (7.69\%) & 17 (65.38\%) & 14 (53.85\%) & 4 (15.38\%) \\
			Time  & Rhino & 5 (19.23\%) & 2 (7.69\%) & 0 (0\%) & 10 (38.46\%) & 9 (34.62\%) & 0 (0.00\%) & 16 (61.54\%) & 12 (46.15\%) & 1 (3.85\%) \\
			\hline
			Ant   & Time  & 3 (11.11\%) & 3 (11.11\%) & 1 (3.7\%) & 5 (18.52\%) & 3 (11.11\%) & --     & 7 (25.93\%) & 6 (22.22\%) & 1 (3.70\%) \\
			Aspectj & Time  & 2 (7.41\%) & 2 (7.41\%) & --     & 4 (14.81\%) & 4 (14.81\%) & 1 (3.70\%) & 5 (18.52\%) & 6 (22.22\%) & -- \\
			Lang  & Time  & 3 (11.11\%) & 2 (7.41\%) & 0 (0.00\%) & 5 (18.52\%) & 5 (18.52\%) & 1 (3.70\%) & 7 (25.93\%) & 6 (22.22\%) & 1 (3.70\%) \\
			Lucene & Time  & 3 (11.11\%) & 2 (7.41\%) & 1 (3.70\%) & 5 (18.52\%) & 4 (14.81\%) & 1 (3.70\%) & 7 (25.93\%) & 7 (25.93\%) & 1 (3.70\%) \\
			Math  & Time  & 3 (11.11\%) & 2 (7.41\%) & 1 (3.70\%) & 5 (18.52\%) & 4 (14.81\%) & 3 (11.11\%) & 7 (25.93\%) & 5 (18.52\%) & 5 (18.52\%) \\
			Rhino & Time  & 3 (11.11\%) & 2 (7.41\%) & 1 (3.70\%) & 6 (22.22\%) & 5 (18.52\%) & 2 (7.41\%) & 8 (29.63\%) & 7 (25.93\%) & 2 (7.41\%) \\
			\hline
		\end{tabular}%
	\end{adjustbox}
	\label{tab:cross_proj_topN}%
\end{table*}%

\begin{table}[t!]	
	\caption{Mean Average Precision (MAP) results in cross-project setting.}
	\centering
	\begin{tabular}{|l|l|c|c|c|}
		\hline
		\multirow{2}[4]{*}{\textbf{Source}} & \multirow{2}[4]{*}{\textbf{Target}} & \multicolumn{3}{c|}{\textbf{MAP}} \\
		\cline{3-5}          &       & \textbf{NetML} & \textbf{AML} & \textbf{Savant} \\
		\hline
		\hline
		Aspectj & Ant   & 0.191 & 0.181 & -- \\
		Lang  & Ant   & 0.185 & 0.188 & 0.054 \\
		Lucene & Ant   & 0.183 & 0.210  & 0.082 \\
		Math  & Ant   & 0.198 & 0.192 & 0.057 \\
		Rhino & Ant   & 0.186 & 0.188 & 0.077 \\
		Time  & Ant   & 0.187 & 0.164 & 0.022 \\
		\hline
		Ant   & Aspectj & 0.106 & 0.08  & \multirow{6}[2]{*}{--} \\
		Lang  & Aspectj & 0.098 & 0.088 &  \\
		Lucene & Aspectj & 0.098 & 0.087 &  \\
		Math  & Aspectj & 0.052 & 0.091 &  \\
		Rhino & Aspectj & 0.103 & 0.100   &  \\
		Time  & Aspectj & 0.156 & 0.066 &  \\
		\hline
		Ant   & Lang  & 0.276 & 0.275 & 0.336 \\
		Aspectj & Lang  & 0.324 & 0.334 & -- \\
		Lucene & Lang  & 0.349 & 0.319 & 0.331 \\
		Math  & Lang  & 0.330  & 0.300   & 0.514 \\
		Rhino & Lang  & 0.330  & 0.260  & 0.387 \\
		Time  & Lang  & 0.335 & 0.296 & 0.491 \\
		\hline
		Aspectj & Lucene & 0.185 & 0.152 & -- \\
		Ant   & Lucene & 0.218 & 0.150  & 0.069 \\
		Lang  & Lucene & 0.228 & 0.150 & 0.044 \\
		Math  & Lucene & 0.166 & 0.150  & 0.07 \\
		Rhino & Lucene & 0.228 & 0.149 & 0.144 \\
		Time  & Lucene & 0.220  & 0.131 & 0.012 \\
		\hline
		Ant   & Math  & 0.236 & 0.235 & 0.075 \\
		Aspectj & Math  & 0.221 & 0.234 & -- \\
		Lang  & Math  & 0.227 & 0.205 & 0.103 \\
		Lucene & Math  & 0.236 & 0.194 & 0.095 \\
		Rhino & Math  & 0.232 & 0.204 & 0.143 \\
		Time  & Math  & 0.208 & 0.202 & 0.195 \\
		\hline
		Ant   & Rhino & 0.188 & 0.148 & 0.023 \\
		Aspectj & Rhino & 0.205 & 0.167 & -- \\
		Lang  & Rhino & 0.177 & 0.126 & 0.095 \\
		Lucene & Rhino & 0.187 & 0.117 & 0.028 \\
		Math  & Rhino & 0.196 & 0.139 & 0.095 \\
		Time  & Rhino & 0.171 & 0.121 & 0.006 \\
		\hline
		Ant   & Time  & 0.109 & 0.096 & 0.047 \\
		Aspectj & Time  & 0.153 & 0.067 & 0.033 \\
		Lang  & Time  & 0.109 & 0.071 & -- \\
		Lucene & Time  & 0.189 & 0.069 & 0.047 \\
		Math  & Time  & 0.109 & 0.067 & 0.104 \\
		Rhino & Time  & 0.171 & 0.075 & 0.081 \\
		\hline
	\end{tabular}%
	\label{tab:cross_proj_MAP}%
\end{table}%

\subsection{Multi-Modal Feature Location} 

Multi-modal feature location takes as input a feature description and a program spectra, and finds program elements that implement the corresponding feature. There are several multi-modal feature location techniques proposed in the literature~\cite{PoshyvanykGMAR07,LiuMPR07,DitRP13}.

Poshyvanyk et al. proposed an approach named PROMESIR that computes weighted sums of scores returned by an IR-based feature location solution (LSI~\cite{MarcusM03}) and a spectrum-based solution (Tarantula~\cite{JH05}), and rank program elements based on their corresponding weighted sums~\cite{PoshyvanykGMAR07}. Then, Liu et al. proposed an approach named SITIR which filters program elements returned by an IR-based feature location solution (LSI~\cite{MarcusM03}) if they are not executed in a failing execution trace~\cite{LiuMPR07}. Later, Dit et al. used HITS, a popular algorithm that ranks the importance of nodes in a graph, to filter program elements returned by SITIR~\cite{DitRP13}.
Several variants are described in their paper and the best performing ones are $IR_{LSI}Dyn_{bin}WM_{HITS}(h,bin)^{bottom}$ and  $IR_{LSI}Dyn_{bin}WM_{HITS}(h,freq)^{bottom}$. We refer to these two as DIT$^A$ and DIT$^B$, respectively. They have showed that these variants outperform SITIR, though they have never been compared with PROMESIR.

In this work, we compare our proposed approach against PROMESIR, DIT$^A$ and DIT$^B$. We show that our approach outperforms all of them on all datasets.

\subsection{IR-Based Bug Localization} 

Various IR-based bug localization approaches that employ information retrieval techniques to calculate the similarity between a bug report and a program element (e.g., a method or a source code file) have been proposed~\cite{Rao:2011:RSL:1985441.1985451,Lukins:2010:BLU:1824820.1824850,LeWL13,Sisman:2012:IVH:2664446.2664454,Zhou:2012:BFM:2337223.2337226,SahaLKP13,WL14,WangLL14,YeBL14}. 

Lukins et al. used a topic modeling algorithm named Latent Dirichlet Allocation (LDA) for bug localization~\cite{Lukins:2010:BLU:1824820.1824850}. Then, Rao and Kak evaluated the use of many standard IR methods for bug localization including VSM and Smoothed Unigram Model (SUM)~\cite{Rao:2011:RSL:1985441.1985451}. In the IR community, VSM has a long history, proposed four decades ago by Salton et al.~\cite{SaltonWY75} and followed by many other IR methods including SUM and LDA, which address the limitations of VSM.


More recently, a number of approaches which consider information aside from text in bug reports to better locate bugs were proposed. Sisman and Kak proposed a version history-aware bug localization method that considers past buggy files to predict the likelihood of a file to be buggy and uses this likelihood along with VSM to localize bugs~\cite{Sisman:2012:IVH:2664446.2664454}. Around the same time, Zhou et al.~\cite{Zhou:2012:BFM:2337223.2337226} proposed an approach named BugLocator that includes a specialized VSM (named rVSM) and considers the similarities among bug reports to localize bugs. Next, Saha et al.~\cite{SahaLKP13} developed an approach that considers the structure of source code files and bug reports and employs structured retrieval for bug localization, and it outperforms BugLocator. Wang and Lo proposed an approach that integrates the approaches by Sisman and Kak, Zhou et al. and Saha et al. for more effective bug localization~\cite{WL14}. Most recently, Ye et al. devised an approach named LR that combines multiple ranking features using learning-to-rank to localize bugs, and these features include surface lexical similarity, API-enriched lexical similarity, collaborative filtering, class name similarity, bug fix recency, and bug fix frequency~\cite{YeBL14}.



All these approaches can be used as the AML$^\text{Text}$ component of our approach. In this work, we experiment with a basic IR technique namely VSM. Our goal is to show that even with the most basic IR-based bug localization component, we can outperform existing approaches including the state-of-the-art IR-based approach by Ye et al.~\cite{YeBL14}.

\subsection{Spectrum-Based Bug Localization} 

Various spectrum-based bug localization approaches have been proposed~\cite{JH05,Abreu:2009,LuciaLJB10,LuciaLJTB14, Libl+05,LYFHM05,Artzi2010,Artzi:2010:PFL:1806799.1806840,ZH02,Zeller2002a,CZ05,LuciaLX14}. These approaches analyze a program spectra which is a record of program elements that are executed in failed and successful executions, and generate a ranked list of program elements. Many of these approaches propose various formulas that can be used to compute the suspiciousness of a program element given the number of times it appears in failing and successful executions.

Jones and Harrold proposed Tarantula that uses a suspiciousness score formula to rank program elements~\cite{JH05}. Later, Abreu et al. proposed another suspiciousness formula called Ochiai~\cite{Abreu:2009}, which outperforms Tarantula. Then, Lucia et al. investigated 40 different association measures and highlighted that some of them including Klosgen and Information Gain are promising for spectrum-based bug localization~\cite{LuciaLJB10,LuciaLJTB14}. Recently, Xie et al. conducted a theoretical analysis and found that several families of suspiciousness score formulas outperform other families~\cite{XieCKX13}. Next, Yoo proposed to use genetic programming to generate new suspiciousness score formulas that can perform better than many human designed formulas~\cite{Yoo12}. Subsequently, Xie et al. theoretically compared the performance of the formulas produced by genetic programming and identified the best performing ones~\cite{XieKCYH13}. Most recently, Xuan and Monperrus combined 25 different suspiciousness score formulas into a composite formula using their proposed algorithm named MULTRIC, which performs its task by making use of an off-the-shelf learning-to-rank algorithm named RankBoost~\cite{XuanM14}. MULTRIC has been shown to outperform the best performing formulas studied by Xie et al.~\cite{XieCKX13} and the best performing formula constructed by genetic programming~\cite{Yoo12,XieKCYH13}.

Wong et al.~\cite{wong2016survey} provided a comprehensive literature review of a large number of spectrum-fault localization techniques, and pointed out avenues for future work. 
Perez et al.~\cite{Perez:2017:TDM:3097368.3097446} proposed DUU, a new metric for evaluating the diagnosability of a test-suite when applying spectrum-based fault localization approaches. 
\rechecknewagain{Sohn et al.~\cite{Sohn:2017:FUC:3092703.3092717} presented FLUCCs, a fault localization technique that learns to rank program elements based on existing spectrum-fault localization techniques and source code metrics such as age, code churn, and complexity. Li et al.~\cite{Li:2017:TPT:3152284.3133916} proposed TraPT, another learning-to-rank approach that transforms programs and test outputs/messages in order to localize faults effectively. 
}
Pearson et al.~\cite{PearsonCJFAEPK2017} highlighted that results found by evaluating spectrum-based and mutation-based fault localization techniques on artificial faults are significantly different than when they are evaluated on real faults. They thus recommended that fault localization techniques should be evaluated using real faults. Moreover, they introduced several new variants of a mutation-based fault localization technique that also use coverage information (in addition to mutation information). The best variant outperforms Dstar by 2-6\% and 3-6\% considering the Top 5 and Top 10 results respectively.

Many of the above mentioned approaches that compute suspiciousness scores of program elements can be used in the AML$^\text{Spectra}$ component of our proposed approach. In this work, we experiment with a popular spectrum-based fault localization technique namely Tarantula, published a decade ago, which is also used by PROMESIR~\cite{PoshyvanykGMAR07}. Our goal is to show that even with a basic spectrum-based bug localization component, we can outperform existing approaches including the state-of-the-art spectrum-based approaches.

\subsection{Other Related Studies} 

There are many studies that compose multiple methods together to achieve better performance. For example, Kocaguneli et al.~\cite{kocaguneli2012value} combined several single software effort estimation models to create more powerful multi-model ensembles. Also, Rahman et al.~\cite{rahman2014comparing} used static bug-finding to improve the performance of statistical defect prediction (and vice versa). Le et al.~\cite{le2015synergizing} proposed SpecForge that combines different automaton based specification miners using  model fission and model fusion in order to create a more effective specification miner.
Kellogg et al.~\cite{Kellogg16} presents \textit{N-Prog} that combines static bug detection and test case generation to avoid unnecessary human effort. In particular, \textit{N-Prog} produces no false alarms, by construction, since its output alarm is either a new test case or a bug in a program.


\section{Conclusion and Future Work}
\label{sec.conclusion}

In this paper, we put forward a novel multi-modal bug localization approach  named \underline{Net}work-clustered \underline{M}ulti-modal Bug \underline{L}ocalization (NetML). Deviating from the contemporary multi-modal localization approaches, NetML is able to achieve an effective bug localization through the interplay of two sets of model parameters characterizing both bug reports and methods. It also features an adaptive learning procedure that stems from a strictly convex objective function formulation, thereby provides a sound theoretical guarantee on the uniqueness of the optimal solution.

We have extensively evaluated NetML on 355 real bugs from seven different software projects (i.e., Ant, AspectJ, Lang, Lucene, Math, Rhino, and Time). Among the 355 bugs, NetML is able to successfully localize 116, 219, and 255 bugs when developers inspect the Top 1, Top 5, and Top 10 methods, respectively. Compared to the best performing baseline (i.e., AML), NetML can successfully localize 31.82\%, 22.35\%, and 19.72\% more bugs when developers inspect the Top 1, Top 5, and Top 10 methods, respectively. Furthermore, in terms of MAP, NetML outperforms the other baselines by 19.24\%. Based on the Wilcoxon signed-rank test using BH procedure, we show that the results of NetML are significantly better across the seven projects, in terms of Top 1, Top 5, Top 10, and MAP scores.

Although NetML offers a powerful bug localization approach, there remains room for improvement. For example, the current approach as well as the IR-based techniques capture both bug reports and program elements (method) using a simple bag-of-words (e.g., TF-IDF) representation, ignoring the inherent structure within the source codes of a program, such as function call and/or data dependencies. 
In the future, we wish to consider a richer set of structural information within a program element, which carries additional semantics beyond the lexical terms. In particular, we would like to leverage both program structure information and lexical source code to localize potential bugs. We also plan to develop a more sophisticated technique, e.g., based on deep learning~\cite{Goodfellow2016}, to automatically learn the feature representation of bug reports and program elements.

\vspace{5mm}\noindent{\bf Dataset and Codes.} The codes and data for NetML are now available at \url{https://github.com/JHoangSMU/NetML}.



%

\appendix

\rechecknewagain{Tables~\ref{tab:full_result_top_N} and~\ref{tab:full_result_top_N_}
provide a detailed breakdown of the Top N results in Tables~\ref{tab:result_top_N} and~\ref{tab:result_top_N_} produced by NetML and the other baseline methods for each software project (i.e., the within-project setting).}

\rechecknewagain{
Meanwhile, Tables~\ref{tab:cross_proj_topN} and~\ref{tab:cross_proj_MAP} present the detailed breakdown of the Top N and MAP results in Table~\ref{tab:cross_proj_sum} for each pair of source and target projects in cross-project setting, respectively.}

\balance

\let\oldthebibliography=\thebibliography
\let\endoldthebibliography=\endthebibliography
\renewenvironment{thebibliography}[1]{%
  \vspace{0pt}
  \begin{oldthebibliography}{#1}%
    \setlength{\parskip}{0ex}%
    \setlength{\itemsep}{0.5ex}%
}%
{%
  \end{oldthebibliography}%
}
{\scriptsize
\def\baselinestretch{0.95}
\bibliographystyle{IEEEtranS}

\bibliography{main,pbuglocator,faultlocal,IEEEabrv}
}

\begin{figure*}
	\begin{subfigure}{1.0\textwidth}
		\begin{minipage}[!t]{0.5\columnwidth}			
			\begin{tabular}{p{1.0\columnwidth}}
				\begin{IEEEbiography}[{\includegraphics[width=1in,height=1.25in,clip,keepaspectratio]{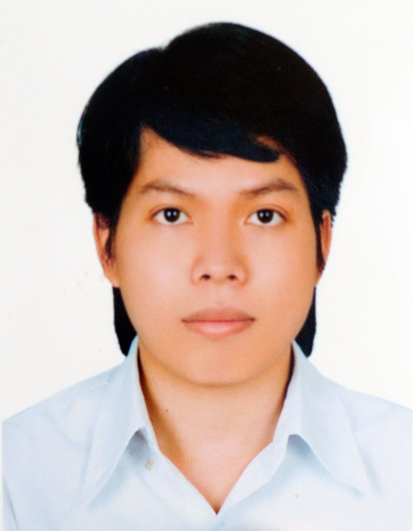}}]%
					{Thong Hoang} is a Ph.D candidate at School of Information Systems, Singapore Management University advised by Associate Professor David Lo. He received his B.Eng and M.S in Computer Science from University of Technology - Vietnam and Konkuk University - South Korea, respectively. He worked as research engineer in three years in Ghent University. He spent for more than a year in School of Information Systems, Singapore Management University. His research interests are software bug localization, defect prediction, deep learning and data mining area. 
				\end{IEEEbiography}
			\end{tabular}
		\end{minipage} \hfill
	\end{subfigure} 
	\begin{subfigure}{1.0\textwidth}
		\begin{minipage}[!t]{0.5\columnwidth}			
			\begin{tabular}{p{1.0\columnwidth}}
				\begin{IEEEbiography}[{\includegraphics[width=1in,height=1.25in,clip,keepaspectratio]{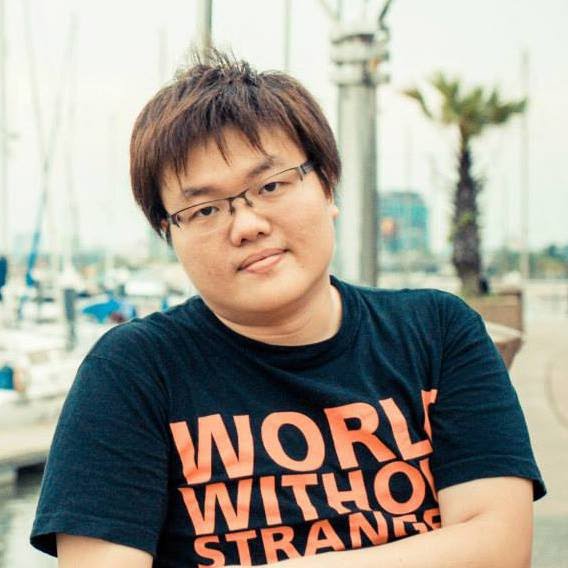}}]%
					{Richard J. Oentaryo} is a Senior Data Scientist at McLaren Applied Technologies Singapore, currently serving as a technical lead of the Public Transport Business Unit in Singapore. Previously, he was a Research Scientist at the Living Analytics Research Centre (LARC), School of Information Systems, Singapore Management University (SMU) in 2011-2016, and a Research Fellow at the School of Electrical and Electronic Engineering, Nanyang Technological University (NTU) in 2010-2011. He received Ph.D. and B.Eng. (First Class Honour) degrees from the School of Computer Engineering (SCE), NTU, in 2011 and 2004 respectively. Dr. Oentaryo is a member of the Association for Computing Machinery (ACM) and Institute of Electrical and Electronics Engineers (IEEE). He has published in numerous international journals and conferences, and received such awards as the IES Prestigious Engineering Achievement Award in 2011, IEEE-CIS Outstanding Student Paper Travel Grant in 2006 and 2009, and ITMA Gold Medal cum Book Prize in 2004.					
				\end{IEEEbiography}
			\end{tabular}
		\end{minipage}
	\end{subfigure}
	\begin{subfigure}{1.0\textwidth}
		\begin{minipage}[!t]{0.5\columnwidth}			
			\begin{tabular}{p{1.0\columnwidth}}
				\begin{IEEEbiography}[{\includegraphics[width=1in,height=1.25in,clip,keepaspectratio]{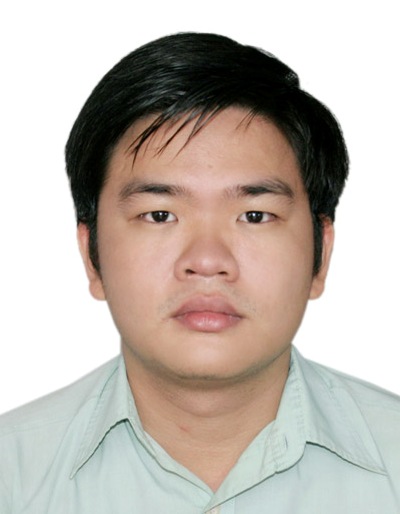}}]%
					{Tien-Duy B. Le} is a Research Scientist at School of Information Systems, Singapore Management University. He received his Ph.D. in Information Systems from Singapore Management University in 2017 under the supervision of Associate Professor David Lo. Before that, he studied and obtained his B.Eng. in Computer Science from Hochiminh City University of Technology, Vietnam in 2012. His research interests include software bug localization and specification mining.				
				\end{IEEEbiography}
			\end{tabular}
		\end{minipage}
	\end{subfigure}
	\begin{subfigure}{1.0\textwidth}
		\begin{minipage}[!t]{0.5\columnwidth}			
			\begin{tabular}{p{1.0\columnwidth}}
				\begin{IEEEbiography}[{\includegraphics[width=1in,height=1.25in,clip,keepaspectratio]{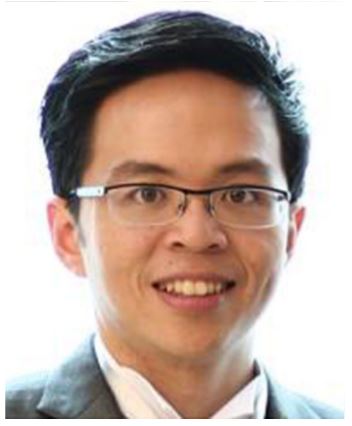}}]%
					{David Lo} received the PhD degree from the School of Computing, National University of Singapore in 2008. He is currently an Associate Professor in the School of Information Systems, Singapore Management University. He has more than
					10 years of experience in software engineering and data mining research and has more than 150 publications in these areas. He has received a number of research awards including two ACM Distinguished Paper awards. He has served in many program and organizing committees, including having served as the general chair of the 31st IEEE/ACM International Conference on Automated Software Engineering (ASE 2016). 			
				\end{IEEEbiography}
			\end{tabular}
		\end{minipage}
	\end{subfigure}
\end{figure*}

\end{document}